\documentstyle[12pt,twoside]{article}

\newcommand{\beqn}{\begin{eqnarray}}
\newcommand{\eeqn}{\end{eqnarray}}
\newcommand{\be}{\begin{equation}}
\newcommand{\ee}{\end{equation}}
\newcommand{\ba}{\begin{array}}
\newcommand{\ea}{\end{array}}
\newcommand{\R}{{\rm\bf R}}
\newcommand{\C}{{\rm\bf C}}
\newcommand{\pa}{\partial}
\newcommand{\re}{\ref}
\newcommand{\ci}{\cite}
\newcommand{\la}{\label}
\newcommand{\bfr}{\begin{flushright}}
\newcommand{\efr}{\end{flushright}}
\newcommand{\bfl}{\begin{flushleft}}
\newcommand{\efl}{\end{flushleft}}
\newcommand{\fr}{\frac}
\newcommand{\ov}{\overline}

\newcommand{\si}{\sigma}
\newcommand{\Si}{\Sigma}
\newcommand{\al}{\alpha}

\newcommand{\ds}{\displaystyle}
\newcommand{\ve}{\varepsilon}

\newcommand{\de}{\delta}
\newcommand{\De}{\Delta}
\newcommand{\na}{\nabla}
\newcommand{\ga}{\gamma}
\newcommand{\om}{\omega}
\newcommand{\Om}{\Omega}

\newcommand{\Lam}{\Lambda}
\newcommand{\br}{|\kern-.25em|\kern-.25em|}
\newcommand{\brr}{{|\kern-.15em|\kern-.15em|\kern-.15em}\,}
\newcommand{\toHe}
{\buildrel {\hspace{2mm}{\cal H}^{-\ve }}\over
{- \hspace{-2mm} \rightharpoondown }}

\begin{document}

 \renewcommand{\theequation}{\thesection.\arabic{equation}}
\newtheorem{theorem}{Theorem}[section]
\renewcommand{\thetheorem}{\arabic{section}.\arabic{theorem}}
\newtheorem{definition}[theorem]{Definition}
\newtheorem{deflem}[theorem]{Definition and Lemma}
\newtheorem{lemma}[theorem]{Lemma}
\newtheorem{example}[theorem]{Example}
\newtheorem{remark}[theorem]{Remark}
\newtheorem{remarks}[theorem]{Remarks}
\newtheorem{cor}[theorem]{Corollary}
\newtheorem{pro}[theorem]{Proposition}
\mathsurround=2pt
\newcommand{\HH}{{\rm\bf H}}
\newcommand{\pr}{\prime}
\def\N{{\rm I\kern-.1567em N}}                              
\def\No{\N_0}                                   
\def\R{{\rm I\kern-.1567em R}}                              
\def\C{{\rm C\kern-4.7pt                                    
\vrule height 7.7pt width 0.4pt depth -0.5pt \phantom {.}}\,}
\def\Z{{\sf Z\kern-4.5pt Z}}                                
\def\n#1{\vert #1 \vert}                   
\def\nn#1{\Vert #1 \Vert}                    
\def\Re {{\rm Re\, }}                                       
\def\Im {{\rm Im\,}}                                        
\newcommand{\loc}{\scriptsize loc}
\newcommand{\const}{\mathop{\rm const}\nolimits}
\newcommand{\sgn}{\mathop{\rm sgn}\nolimits}
\newcommand{\supp}{\mathop{\rm supp}\nolimits}
\newcommand{\mod}{\mathop{\rm mod}\nolimits}
\newcommand{\ow}{\overrightarrow}

\begin{titlepage}
\hspace{5cm}  {\it Markov
 Processes and Related Fields}  {\bf 8} (2002), no.1, 43-80
\vspace{2cm}

\hspace{4cm} {\it Dedicated to M.I.Vishik on the occasion of his 80th 
anniversary}
\vspace{0.5cm} 
\begin{center}
 {\Large\bf  On  a Two-Temperature Problem
 for Wave Equation }\\
 \bigskip

{\large T.V.~Dudnikova}
\footnote{Supported partly by
research grants of DFG (436 RUS 113/615/0-1) and of RFBR (01-01-04002).}\\
{\it  Mathematics Department\\
        Elektrostal Polytechnic Institute\\
        Elektrostal 144000, Russia }\\
e-mail:~misis@elsite.ru
 \bigskip\\
 {\large A.I.~Komech}
\footnote{Supported partly by 
CONACYT (Mexico), by the Institute of Physics and Mathematics
of the University of Michoacan (Morelia, Mexico),
Max-Planck Institute for Mathematics in the Sciences
(Leipzig, Germany),  
research grants of DFG (436 RUS 113/615/0-1)
 and of RFBR (01-01-04002).}\\
 {\it Mechanics and Mathematics Department \\
Moscow State University\\
 Moscow 119899, Russia}\\
 e-mail:~komech@mech.math.msu.su
 \bigskip\\
 {\large H.~Spohn}\\
{\it Zentrum Mathematik\\
Technische Universit\"at\\
M\"unchen D-80290, Germany}\\
 e-mail:~spohn@mathematik.tu-muenchen.de\\
 \end{center}
 \vspace{0.5cm}

 \begin{abstract}
 Consider the  wave equation
with constant or variable coefficients
 in   $\R^3$.
The initial datum  is a  random function
with a finite mean density of energy
that also satisfies  a Rosenblatt- or
Ibragimov-Linnik-type mixing condition.
The random function
converges to different space-homogeneous
processes as $x_3\to\pm\infty$, with the distributions  $\mu_\pm$.
We study the distribution $\mu_t$
 of the random solution at a time $t\in\R$.
The main result is the convergence of $\mu_t$ to
 a Gaussian translation-invariant measure as $t\to\infty$
that means central limit theorem
for the wave equation.
The proof
 is based on
the  Bernstein `room-corridor'  argument.
The application to the case of the Gibbs measures
$\mu_\pm=g_\pm$ with two different
temperatures $T_{\pm}$  is given.
Limiting mean energy current density {\it formally}
is $-\infty\cdot (0,0,T_+\!\!-\!T_-)$
for the Gibbs measures,
and it is finite and equals to
$-C(0,0,T_+\!\!-\!T_-)$ with $C>0$
for the convolution with a nontrivial test function.
 \end{abstract}
\end{titlepage}

\section{Introduction}
The paper concerns a mathematical problem of foundations of
statistical physics.
We consider the
 Second Law of thermodynamics in a reversible
infinite dimensional Hamiltonian equations.
The Law states that the energy current is
 directed from higher  temperature to lower temperature.
We derive the Law for wave equations in $\R^3$
 with constant and  variable coefficients.
The key role  plays
the mixing condition of Rosenblatt- or Ibragimov-Linnik-type
for an initial measure.
The mixing condition is introduced initially
by R.L.~Dobrushin and Ya.M.~Suhov
in their approach
to the problem
of foundation of statistical physics
for infinite-particle systems, \ci{DS1, DS2}.
The mixing condition is used also in
the paper
\ci{BPT}
which concerns
 a discrete version of our result for a 1D
chain of harmonic oscillators.
Let us explain our
 result in the case of constant coefficients,
\beqn\la{1}
\left\{
\ba{l}
\ddot u(x,t)=   \De u(x,t),\,\,\,\,x\in\R^3,\\
~\\
u\Bigl|_{t= 0}= u_0(x),
~~\dot u\Bigl|_{t= 0}= v_0(x).
\ea
\right.
\eeqn
Denote
$
Y(t)=(Y^0(t),Y^1(t))\equiv (u(\cdot,t),\dot u(\cdot,t))$,
$Y_0=(Y^0_0,Y^1_0)\equiv (u_0,v_0)$.
Then  (\ref{1}) becomes
\be\la{1'}
\dot Y(t)={\cal F}(Y(t)),\,\,\,t\in\R,\,\,\,\,Y(0)=Y_0.
\ee
 We assume that
the initial datum $Y_0$ is a
random function with zero mean living in a
 functional phase space ${\cal H}$ of states of finite local energy;
the distribution of $Y_0$ is denoted by $\mu_0$.
Denote by  $\mu_t(dY)$, $t\in\R$, the measure on  ${\cal H}$
giving the distribution of
the random solution  $Y(t)$
to  problem (\re{1'}).
We assume that the initial
correlation functions
$Q_0^{ij}(x,y)\equiv E \Big(Y_0^i(x)Y_0^j(y)\Big)$, $i,j=0,1$,
and some of their derivatives
are continuous
and decaying as $|x-y|\to\infty$.
In particular, the
initial mean energy density  is bounded:
\be\la{med}
 E [\vert \nabla u_0(x)\vert^2
 + \vert v_0(x)\vert^2]
=[\na_x\cdot\na_y Q_0^{00}(x,y)]|_{y=x}+Q_0^{11}(x,x)
\le C
<\infty,\,\,\,x\in\R^3.
\ee
Next,
we assume that
the initial
correlation matrix $(Q_0^{ij}(x,y))_{i,j= 0,1}$
has the  form
\beqn\la{3}
Q_0^{ij}(x,y)
= \left\{
\ba{l}
q_-^{ij}(x-y),~~x_3, y_3<\!\!-a,\\
~\\
q_+^{ij}(x-y),~~x_3, y_3>\,a.\ea
\right.
\eeqn
Here $q_{\pm}^{ij}(x-y)$ are
 the correlation functions of some
translation-invariant measures $\mu_{\pm}$
with zero mean value  in ${\cal H}$,
$x= (x_1,x_2,x_3),~ y= (y_1,y_2,y_3)\in\R^3$,
and $a>0$.
The measure  $\mu_0$ is not translation-invariant
if $q_-^{ij}\not=q_+^{ij}$.
Finally, we assume that the initial measure $\mu_0$ satisfies
a mixing
condition.
Roughly speaking, it means that
\be\la{mix}
Y_0(x)\,\,\,\,   and \, \, \,\,Y_0(y)
\,\,\,\,  are\,\,\,\, asymptotically\,\,\,\, independent\,\, \,\,
 as \,\, \,\,
|x-y|\to\infty.
\ee
Our main result establishes  the (weak)
convergence
\be\la{1.8i}
\mu_t \rightharpoondown
\mu_\infty,\,\,\,\, t\to \infty,
\ee
to an
 equilibrium measure $\mu_\infty$,
which is a translation-invariant Gaussian measure on ${\cal H}$.
 The similar convergence holds for  $t\to-\infty$
since our system is time-reversible.
We construct generic examples of the random initial datum satisfying
all assumptions imposed.
We get the explicit formulas (\re{q00})-(\re{q11})
for the limiting
 correlation matrices.
\medskip

We apply our results to the case of the Gibbs measures
$\mu_\pm=g_{\pm}$.
Formally
\be\la{5}
g_{\pm}(du_0, dv_0)= \frac{1}{Z_\pm}
\ds e^{-\ds  \fr {\beta_{\pm}}2\ds\int(|\nabla u_0(x)|^2+|v_0(x)|^2)
dx}\prod_{x}
du_0(x)dv_0(x),~~\beta_{\pm}= T^{-1}_{\pm},
\ee
where $T_\pm\ge0$ are the corresponding absolute temperatures.
We adjust the definition of the Gibbs measures $g_{\pm}$
in Section 3.
The Gibbs measures $g_{\pm}$ have singular correlation
functions and do not satisfy
our assumptions (\ref{H2}).
Respectively,  our results can not be applied
directly to  $g_{\pm}$. We reduce the problem
by a convolution with a smooth function
$\theta\in D\equiv C_0^\infty(\R^3)$:
we consider Gaussian processes $u_\pm$ corresponding to the
measures $g_{\pm}$ and  define
the ``smoothened'' measures $g^\theta_\pm$
as the distributions of the convolutions $u_\pm*\theta$.
The  measures $g^\theta_\pm$ satisfy all our assumptions,
and  the convergence $g^\theta_t\rightharpoondown g^\theta_\infty$
follows
from (\ref{1.8i}).
 This implies the weak convergence of the measures
$g_t\rightharpoondown g_\infty$
since
$\theta$ is arbitrary.
We show that the limit energy current for $g_\infty$ is formally
$$
\ov j_\infty=-\infty\cdot (0,0,T_+-T_-).
$$
The infinity
 means the ``ultraviolet divergence''.
This  relation is meaningful in the case of smoothened measures
 $g^\theta_\infty$,
$$
\ov j^\theta_\infty=-C_\theta\cdot(0,0,T_+-T_-),
$$
if $\theta(x)$ is axially symmetric with respect to $Ox_3$;
$C_\theta >0$ if $\theta(x)\not\equiv 0$.
This corresponds to the Second Law of thermodynamics.
\medskip

We prove the convergence (\re{1.8i}) in three steps using the strategy of
\ci{DKR,R1,R2}.\\
{\bf I.} The family of measures
 $\mu_t$, $t\geq 0$, is weakly
compact in an appropriate Fr\'echet space.\\
{\bf II.} The correlation functions converge to a limit: for $i,j=0,1,$
 \be\la{corf}
Q_t^{ij}(x,y)=\int Y^i(x)Y^j(y) \mu_t(dY)
\to Q_\infty^{ij}(x,y),\,\,\,\,t\to\infty.
\ee
{\bf III.}
The characteristic functionals converge to the Gaussian:
\be\la{2.6i}
 \hat\mu_t(\Psi )
 =   \int \exp({i\langle Y,\Psi\rangle })\mu_t(dY)
\rightarrow \ds \exp\{-\fr{1}{2}{\cal Q}_\infty (\Psi , \Psi)\},
\,\,\,\,t\to\infty,
\ee
where
$\Psi$ is an arbitrary element of the dual space,  and
${\cal Q}_\infty$ is the quadratic form
with the integral kernel
$(Q^{ij}_\infty(x,y))_{i,j=0,1}$.
\medskip

Property {\bf I} follows from the Prokhorov
Compactness Theorem by using
 methods of  \ci{VF}. First, one proves a uniform bound for the
mean local energy  with respect to the  measure
$\mu_t$.
The conditions of Prokhorov' Theorem then
  follow
from  Sobolev's
Embedding Theorem.
We deduce the uniform bound from the explicit expression
for the correlation functions
$Q_t^{ij}(x,y)$.
 The expression follows from the
Kirchhoff formula for the solutions
to (\re{1}). In particular, in the case $u_0(x)\equiv 0$, we have
\be\la{Kir}
u(x,t)= \frac{1}{4\pi t}\int\limits_{S_t(x)}
v_0(x')  dS(x'),
\ee
where $dS(x')$ is the Lebesgue measure  on the sphere
 $S_t(x):\,|x'-x|=t$.

Property {\bf II} also follows from  explicit formulas
for $Q_t^{ij}(x,y)$.
The formula (\re{Kir}) allows to express the
correlation functions
$Q_t^{ij}(x,y)$
in terms of integrals over spheres of radius $t$.
In the limit, $t\to\infty$, the spheres become the planes.
Respectively,  $Q_\infty^{ij}(x,y)$ is expressed in terms of
integrals of the Radon transform
of initial correlation functions  $Q_0^{ij}(x,y)$.
We reduce the expressions
to some convolutions.
\medskip\\
{\bf Remarks} i) The dynamics (\re{1})
is translation invariant,
and its Fourier transform
has  a very simple form.
 However,
 the proof of (\re{corf})
in Fourier transform
is not transparent
and requires additional efforts
since our main assumption (\re{3})
is stated in the coordinate space.\\
ii)
Our proof of the convergence (\re{corf}) in Sections
5 and 6 does not allow a simplification
in   the particular case of the Gibbs measures (\re{5}).
This is related to the slow long-range
decay of the correlation function
$Q_0^{00}(x,y)\sim |x-y|^{-1}$, $|x-y|\to\infty$.
\medskip

We deduce property {\bf III}
using the method of \ci{DKR}.
The method  is based on a modification
of the Bernstein room-corridor' argument,
and it is suggested by the structure of the
Kirchhoff formula
(\re{Kir}):
roughly speaking, 
(\re{Kir}) is ``the sum'' of weakly dependent
random values devided by the square root of their ``number''.
This observation allows us to reduce the proof of (\ref{2.6i})
to the Lindeberg Central Limit Theorem, similarly to \ci{DKR}.
We do not consider the case $n=2$: it  requires a different approach
since the 
strong Huyghen's principle breaks down.

Let us note that
 our mixing condition is weaker than that in
\ci{DKR}: 
this is necessary in the application to the Gibbs measures (\re{5}).
Namely, we introduce different
 mixing coefficients for
partial derivatives of the random solution at $t=0$: 
we assume that the long-range
decay of the mixing coefficients depends on the order of the derivatives.
Respectively, our proof
 requires new tools (see Sections 7, 9, 10).
For instance,
the splitting (\ref{I(Sigma)})
and the bound (\ref{112}) play a crucial role.

All the three steps {\bf I}-{\bf III}
 of the argument rely  on the mixing
condition. Simple examples show that the convergence
to a Gaussian measure may fail when the
 the mixing condition fails (see \ci{DKR}).

In conclusion,
 we extend the convergence in (\re{1.8i}) to the equations
with  variable coefficients,
 that are constant outside a finite region.
The extension follows immediately from our result for
constant coefficients, using method of \ci{DKR}. The method
is based on the scattering theory for the solutions of infinite
global energy, which is constructed in \ci{DKR}.

The paper is organized as follows.
In Section 2 we formally state our main result.
We apply it to the
 Gibbs
measure in Section 3.
Sections 4-10 deal with the case of constant coefficients:
the compactness (Property {\bf I}) and
the convergence (\ref{corf}) are proved in Sections 4-6.
In  Section 7 we introduce the `room-corridor' method,
in Section 8 we prove   the convergence (\ref{2.6i}),
 and in
Sections 9, 10 we check the Lindeberg condition.
In Section 11 we establish the convergence (\ref{1.8i})
for variable coefficients.
Appendix A concerns the
Radon transform and convolution, and
Appendix B concerns  the Gaussian measures in the
weighted Sobolev spaces.

\bigskip

Let us note that
the equation (\re{1})
describes a continuous $n$-dimensional
family of harmonic oscillators.
Therefore, our result is an extension of the results
\cite{BPT,SL} that concern the infinite one-dimensional
 chains of harmonic oscillators.

Our formulas for the limit correlation functions correspond to the
 discrete one-dimensional version \ci{BPT}.
For instance, the position-momentum correlations
have a power long-range decay. On the other hand, in \ci{RLL, N}
the limit correlation functions  are constructed for the finite chains of $N$
oscillators with the ``Langevin''
boundary value conditions.
In the limit $N\to\infty$
the correlation functions have an exponential long-range decay.
This means that this limit leads to another stationary measure of the infinite
chain, different from \ci{BPT, SL}.

The convergence to statistical equilibrium
 for the wave
equation is established
in \ci{DKR} (see also \ci{R1,R2})
for the case of a translation-invariant
initial measure $\mu_0$.
This corresponds to our result in the
 particular case 
$T_-=T_+$.
The similar result has been proved for the Klein-Gordon equation,
\ci{DKK,K3}.
 If the initial measure
 $\mu_0$ coincides with one of the equilibrium limit measures
 $\mu_\infty$, the corresponding random
solution
$Y(t)$
 is  mixing in time, \ci{D4,KD,D7}.

\setcounter{equation}{0}
\section{Main results}

\subsection{Notations}
We assume that the initial datum $Y_0$
 belongs to the phase space ${\cal H}$ defined below.
 \begin{definition}                 \la{d1.1}
  $ {\cal H} \equiv H_{loc}^1(\R^3)\oplus H_{loc}^0(\R^3)$
 is the Fr\'echet space
 of pairs $Y\equiv(u(x),v(x))$
of real  functions $u(x)$, $v(x)$,
 endowed with  the local energy seminorms
 \beqn                              \la{1.5}
\Vert Y\Vert^2_{R}= \int\limits_{|x|<R}
(|u(x)|^2+|\nabla u(x)|^2+|v(x)|^2) dx<\infty,
~~\forall R>0.
 \eeqn
  \end{definition}

Proposition \re{p1.1} follows from
\ci[Thms V.3.1, V.3.2]{Mikh} as the speed of propagation
for Eqn (\re{1}) is finite.
 \begin{pro}    \la{p1.1}
i) For any $Y_0 \in {\cal H}$
 there exists  a unique solution
$Y(t)\in C(\R, {\cal H})$
 to the Cauchy problem (\re{1'}).\\
 ii) For any $t\in \R$, the operator $U(t):Y_0\mapsto  Y(t)$
 is continuous in ${\cal H}$.\\
iii) The energy inequalities hold $\forall R >0$,
\be\la{een}
\Vert U(t) Y_0\Vert_R\le C(t)\Vert Y_0\Vert_{R+|t|},\,\,\,t\in\R.
\ee
\end{pro}
Let us choose a function $\zeta(x)\in C_0^\infty(\R^3)$ with $\zeta(0)\ne 0$.
Denote by $H^s_{\rm loc}(\R^3),$ $s\in \R,$  the local Sobolev spaces,
i.e. the Fr\'echet spaces
of distributions $u\in D'(\R^3)$ with the finite seminorms
$$
\Vert u\Vert _{s,R}:= \Vert\Lambda^s\Big(\zeta(x/R)u\Big)\Vert_{L^2(\R^3)},
$$
where $\Lambda^s v:=F^{-1}_{k\to x}(\langle  k\rangle^s\hat v(k))$,
$\langle  k\rangle:=\sqrt{|k|^2+1}$, and $\hat v:=F v$ is the Fourier
transform of a tempered distribution $v$.
For $\psi\in D\equiv C_0^\infty(\R^3)$ define
$
F\psi ( k)= \ds\int e^{i k\cdot x} \psi(x) dx.
$

\begin{definition}\la{d1.2}
For $s\in\R$  denote
$
{\cal H}^{s}\equiv H_{\rm loc}^{1+s }(\R^3)
\oplus H_{\rm loc}^{s }(\R^3).
$
\end{definition}

Using the standard techniques of pseudodifferential operators
and Sobolev's Theorem
(see, e.g. \ci{H3}), it is possible to prove that
 ${\cal H}^0={\cal H}\subset {\cal H}^{-\ve }$ for every $\ve>0$,
and the embedding  is compact.
We denote by $\langle \cdot ,\cdot \rangle $
the scalar product in real Hilbert space
$L^2(\R^3)$ or in
$L^2(\R^3)\otimes \R^N$
or in its various extensions.

\subsection{Random solution. Convergence to equilibrium}

Let $(\Om,\si,P)$ be a probability space
 with the expectation $E$, let
 ${\cal B}({\cal H})$ denotes the Borel $\si$-algebra
in ${\cal H}$.
We assume that $Y_0=Y_0(\om,x)$ in (\re{1'})
is a measurable
random function
with values in $({\cal H},\,{\cal B}({\cal H}))$.
In other words,
 $(\om,x)\mapsto Y_0(\om,x)$ is a measurable map
$\Om\times\R^3\to\R^2$ with respect to the
(completed) $\si$-algebra
$\Sigma\times{\cal B}(\R^3)$ and ${\cal B}(\R^2)$.
Then
$Y(t)=U(t) Y_0$ is also a measurable    random function
with values in
$({\cal H},{\cal B}({\cal H}))$, due to Proposition \re{p1.1}.
We denote by $\mu_0(dY_0)$ the Borel probability measure in ${\cal H}$
that is the distribution of  $Y_0$.
Without loss of generality,
 we assume $(\Om,\Si,P)=
({\cal H},{\cal B}({\cal H}),\mu_0)$
and $Y_0(\om,x)=\om(x)$ for
$\mu_0(d\om)\times dx$-almost all
$(\om,x)\in{\cal H}\times\R^3$.

\begin{definition}
$\mu_t$ is the Borel probability measure in ${\cal H}$
that is
the distribution of $Y(t)$:
\begin{eqnarray}\la{1.6}
\mu_t(B) = \mu_0(U(-t)B),\,\,\,\,
\forall B\in {\cal B}({\cal H}),
\,\,\,   t\in \R.
\eeqn
\end{definition}

Our main goal is to derive
 the convergence of the measures $\mu_t$ as $t\rightarrow \infty $.
We establish the weak convergence of  $\mu_t$
in the Fr\'echet spaces ${\cal H}^{-\ve }$ with any  $\ve>0$:
\be\la{1.8}
\mu_t\,\buildrel {\hspace{2mm}{\cal H}^{-\ve }}\over
{- \hspace{-2mm} \rightharpoondown }
\, \mu_\infty
\quad {\rm as}\quad t\to \infty,
\ee
where $\mu_\infty$ is the Borel probability measure in the space
${\cal H}$.
  By definition,    this means
 the convergence
 \be\la{1.8'}
 \int f(Y)\mu_t(dY)\rightarrow
 \int f(Y)\mu_\infty(dY)\quad\rm{as}\quad t\to \infty
 \ee
 for any bounded continuous functional $f(Y)$
 in the space ${\cal H}^{-\ve }$.

\begin{definition}
The correlation functions of the measure $\mu_t$ are
defined by
\be\la{qd}
Q_t^{ij}(x,y)\equiv E \Big(Y^i(x,t)Y^j(y,t)\Big),~~i,j= 0,1,~~~~
{\rm for~~almost~~all}~~~
x,y\in\R^3\times\R^3
\ee
if the expectations in the RHS are finite.
\end{definition}

We set ${\cal D}=D\oplus D$, and
$
\langle Y,\Psi\rangle
=
\langle Y^0,\Psi^0\rangle +\langle Y^1,\Psi^1\rangle
$
for
$Y=(Y^0,Y^1)\in {\cal H}$, and $
\Psi=(\Psi^0,\Psi^1)\in  {\cal D}$.
For a Borel probability  measure $\mu$ in the space ${\cal H}$
we denote by $\hat\mu$
the characteristic functional (Fourier transform)
$$
\hat \mu(\Psi )  \equiv  \int\exp(i\langle Y,\Psi \rangle )\,\mu(dY),\,\,\,
 \Psi\in  {\cal D}.
$$
A  measure $\mu$ is called Gaussian (with zero expectation) if
its characteristic functional has the form
$$
\ds\hat { \mu} (\Psi ) =  \ds \exp\{-\fr{1}{2}
 {\cal Q}(\Psi , \Psi )\},\,\,\,\Psi \in {\cal D},
$$
where ${\cal Q}$ is a  real nonnegative quadratic form in ${\cal D}$.
A measure $\mu$ is called
translation-invariant if
$$
\mu(T_h B)= \mu(B),\,\,\,\,\,\forall B\in{\cal B}({\cal H}),
\,\,\,\, h\in\R^3,
$$
where $T_h Y(x)= Y(x-h)$.

\subsection{Mixing condition}
 Let $O(r)$ denote the set of all pairs of open subsets
 $ {\cal A}, {\cal B} \subset \R^3$ of distance
 $\rho ( {\cal A},\, {\cal B})\geq r$, let
$\al=(\al_1,\al_2,\al_3)$ with integers $\al_i\ge 0$.
Denote by
$\sigma_{i\alpha} ( {\cal A})$
 the $\sigma$-algebra of the subsets in ${\cal H}$ generated by all
  linear functionals
$$
 Y\mapsto \langle D^{\alpha}Y^i,\psi\rangle=\int\limits_{\R^3}
D^{\alpha}Y^i(x)\psi(x)\,dx,\,\,\,|\alpha|\le 1-i,\,\,\,i=0,1,
$$
where
$\psi\in D$
 with $ \supp \psi \subset {\cal A}$.
For $d=0,1$ let
$\sigma_d$ be the $\sigma$-algebra generated by
$\sigma_{i\alpha}$ with $i+|\alpha|\ge d$, i.e.
$$
\sigma_d\equiv \bigvee\limits_{i+|\alpha|\ge d}\sigma_{i\alpha},
\quad d=0,1.
$$
We define the
 Ibragimov-Linnik mixing coefficient
of a probability measure $\mu_0$ on ${\cal H}$
(cf. \ci[Dfn 17.2.2]{IL})
for $d_1,d_2=0,1$ as
$$
   \phi_{d_1,d_2}(r)\equiv
 \sup_{( {\cal A}, {\cal B})\in O(r)} \sup_{
 \ba{c} A\in\si_{d_1}({\cal A}),B\in\si_{d_2}({\cal B})\\
\mu_0(B)>0\ea}
 \fr{| \mu_0(A\cap B) - \mu_0(A)\mu_0(B)|}{ \mu_0(B)}.
$$
\begin{definition}
The measure $\mu_0$ satisfies the strong uniform
 Ibragimov-Linnik mixing condition if for any
$d_1, d_2=0,1$
\be\la{1.11}
\phi_{d_1,d_2}(r)\to 0,\,\,\,r\to\infty.
\ee
\end{definition}

Below we specify  the rate of the decay.

\subsection{Main theorem}
Let  $\nu_d\in C[0,\infty)$ denote some continuous
nonnegative
nonincreasing functions in
$[0,\infty)$ ($d=0,1,2$) with the finite integrals,
\be \la{phi}
\ds\int\limits_0^{\infty}(1+r)^{d-1}\nu_d(r) dr<\infty.
\ee
We also denote $\nu(r)=\nu_2(r)$.
We assume that the measure $\mu_0$ satisfies
the following conditions {\bf S0}-{\bf S3}:
\bigskip\\
{\bf S0}  $\mu_0$ has  the zero expectation value,
\be\la{H0}
EY_0(x)= 0,\quad \,\,x\in\R^3.
\ee
{\bf S1} The correlation functions
of  $\mu_0$ have the form (\re{3}).\\
 ~\\
{\bf S2} The following derivatives are continuous and the bounds hold,
\beqn\la{H2}
|D_{x,y}^{\alpha,\beta} Q^{ij}_{0}(x,y)|\le
\left\{
\ba{l}
C\nu_d(|x-y|)\mbox{   if   }d=0\,\,{\rm or}\,\, 1,\\
C\nu_2(|x-y|)\mbox{   if   }2\le d\le 4,
\ea
\right|\,\,\,\,\,d=i+j+|\alpha|+|\beta|.
\eeqn
{\bf S3} The measure $\mu_0$ satisfies the {\it strong uniform}
 Ibragimov-Linnik mixing condition, and
for $ d_1,d_2=0,1$
\be
   \phi_{d_1,d_2}(r)\le
C\nu_d^2(r),\,\,\,\,\,d=d_1+d_2.
\la{7}
\ee

\begin{remark}\la{rQ}
{\rm i) Condition
{\bf S2} implies (\re{med}). Condition {\bf S3}
implies the estimates (\ref{H2}) with $i+|\alpha|\le 1$, $j+|\beta|\le 1$.\\
ii) The conditions {\bf S2} and {\bf S3}
allow various modifications.
We choose the variant which allow an application to the case of
the Gibbs measures
(\re{5}) (see the next section).
Our mixing condition {\bf S3} is weaker than
the mixing condition
 \ci{DKR} which corresponds to {\bf S3}
with the functions $\nu_{0,1}(r)\le\nu_2(r)$.
On the other hand, the
estimates (\re{H2}) with $d>2$   are not required in \ci{DKR}.
}
\end{remark}

Let ${\cal E}(x)= -\ds\fr 1{4\pi|x|}$
 be the fundamental solution of the Laplacian,
i.e. $\triangle{\cal E}=\delta(x)$ for  $x\in\R^3$, and
 $P(x)=\ds-i
F^{-1}\ds\frac{\sgn k_3}{|k|}
$ where
$F^{-1}$  is
the inverse Fourier transform.
Define, for almost all $x, y\in\R^3$,  the matrix-valued function
 \be\la{Q}
Q_\infty(x,y)=
 \Bigl(Q_\infty ^{ij}(x, y)\Bigr)_{i,j= 0,1} =
 \Bigl(q_\infty ^{ij}(x-y)\Bigr)_{i,j= 0,1}\,\,,
\ee
where
\beqn
q^{00}_{\infty}&=&\ds\frac{1}{4}
\left[
q_+^{00} +q_-^{00}-
 {\cal E}*(q_+^{11} +q_-^{11})+
P*(q_+^{01}-q_-^{01}-q_+^{10}+q_-^{10})\right],\la{q00}\\
q^{10}_{\infty}=\,\,\,-\,\,q^{01}_{\infty}&=&\ds\frac{1}{4}
\left[
q_+^{10} +q_-^{10} -q_+^{01}
-q_-^{01}\,\,\,\,\,\,\,\,\,+
P*(q_+^{11}-q_-^{11}-
\De q_+^{00}+\De q_-^{00})
\right],\la{q10}\\
q^{11}_{\infty}=-\De q^{00}_{\infty}&=&\ds\frac{1}{4}
\left[q_+^{11} +q_-^{11}-
\De (q_+^{00} +q_-^{00})+
P*\De(q_+^{10}-q_-^{10}-
q_+^{01}+q_-^{01})
\right].\la{q11}
\eeqn
 The definition of the convolutions with
$P$
 in formulas
 (\ref{q00})--(\ref{q11})
is adjusted in Appendix A
(formula (\ref{l3.32})).

Denote by
${\cal Q}_{\infty} (\Psi, {\Psi})$ the real quadratic form
in
${\cal D}$   defined by
\be\la{qpp}
{\cal Q}_\infty (\Psi, {\Psi})=\sum\limits_{i,j=0,1}~
\int\limits_{\R^3\times\R^3}
Q_{\infty}^{ij}(x,y)\Psi^i(x)\Psi^j(y)
dx~dy.
\ee
Our main result is the following theorem.
\begin {theorem}\la{t1.2}
Let {\bf S0}-{\bf S3} hold. Then
\\
i) the convergence in (\ref{1.8}) holds for any $\ve>0$. \\
ii)  The limiting measure
$ \mu_\infty $ is a Gaussian equilibrium
measure on ${\cal H}$.\\
iii) The limiting characteristic functional has the form
$$
\ds\hat { \mu}_\infty (\Psi ) =  \ds \exp\{-\fr{1}{2}
{\cal  Q}_\infty( \Psi , \Psi )\},~~ \,\,\, \Psi \in  {\cal D},
$$
where
 ${\cal Q}_\infty $ is the
 quadratic form  with the integral kernel
 $Q_{\infty}(x,y)$ defined in (\ref{Q})-(\ref{q11}).
\end{theorem}

Theorem \ref{t1.2} can be deduced from
 Propositions \re{l2.1} and \re{l2.2} below,
by using the same arguments as in \ci[Thm XII.5.2]{VF}.
  \begin{pro}\la{l2.1} The
 family of the measures $\{\mu_t,\, t\geq 0\}$,
 is weakly compact in
  ${\cal H}^{-\ve }$ with any
$\ve >0$.
 \end{pro}
  \begin{pro}\la{l2.2} For any $\Psi\in  {\cal D}$,
 \be\la{2.6}
 \hat\mu_t(\Psi )\equiv
 \int\exp(i\langle Y,\Psi \rangle )\,\mu_t(dY)
\rightarrow \ds \exp\{-\fr{1}{2}{\cal Q}_\infty (\Psi , \Psi)\},
\,\,\,\,\,\,t\to\infty.
 \ee
 \end{pro}

 Proposition \re{l2.1} is proved in Section 4
for a simple particular case, and in
Section 6 for the general case.
Proposition \re{l2.2} is proved
Sections 7, 8.

\subsection{Examples}
\subsubsection{Gaussian measures}
We construct the
Gaussian initial measures $\mu_0$ satisfying {\bf S0}--{\bf S3}.
Let us take some Gaussian measures $\mu_\pm$ in ${\cal H}$
with  correlation
functions $q_\pm^{ij}(x-y)$ which are zero for
 $i\not= j$, while for $i=0,1$,
\beqn
\left.
\ba{rcl}
q_\pm^{ii}(z)&=&F^{-1}\hat q_\pm^{ii}(k),\\~&&\\
(1+|k|)^{s}
\pa^\ga_k
 \hat q_\pm^{ii}(k)&\in& L^1(\R^3),\,\,\,\,\,0\le d=2i+s\le 4,
\,\,\,\,\,|\ga|\le 1+d,
\\~&&\\
\hat q_\pm^{ii}(k)&\ge& 0.
\ea
\right| \la{S04}
\eeqn
Then $\mu_\pm$ satisfy {\bf S0}, {\bf S2}
with the functions $\nu_d(r)=C(1+r)^{-1-d}$ for a sufficiently
large 
$C>0$.
Let us take the functions $\zeta_{\pm}\in C^{\infty}(\R)$ s.t.
$$
\zeta_{\pm}(s)= \left\{ \ba{ll}
1,~~\mbox{for }~ \pm s>\,a,\\
0,~~\mbox{for }~ \pm s<-a.\ea\right.
$$
Let us introduce
$(Y_-,Y_+)$ as
a unit random function in the probability space
$({\cal H}\times{\cal H}, \mu_-\times\mu_+)$.
Then $Y_{\pm}$ are
 Gaussian independent vectors
in
${\cal H}$.
Define  $\mu_0$ as  the 
distribution of the random function
\be\la{rf}
Y_0(x)= \zeta_-(x_3)Y_-(x)+\zeta_+(x_3)Y_+(x).
\ee
Then correlation functions of  $\mu_0$
are
\be\la{q1}
Q_0^{ij}(x,y)=
q_-^{ij}(x-y)\zeta_-(x_3)\zeta_-(y_3)+
q_+^{ij}(x-y
)\zeta_+(x_3)\zeta_+(y_3),~~i,j= 0,1,
\ee
where $x= (x_1,x_2,x_3)$,
$y= (y_1,y_2,y_3)\in\R^3$,
$q_{\pm}^{ij}$ are the correlation functions  of the measures
$\mu_{\pm}$. Then  {\bf S0} and {\bf S1} hold,
 and {\bf S2} follows for $\mu_0$
with the same functions $\nu_d(r)$ as for $\mu_\pm$.
Let us assume, in addition to (\re{S04}), that
\be\la{S5}
q_\pm^{ii}(x)=0,\,\,\,|x|\geq r_0.
\ee
Then  the mixing condition (\re{1.11})
 holds since $\phi_{d_1,d_2}(r)=0$, $r\geq r_0$,
and {\bf S3} follows.
For instance,   (\re{S04}) and (\re{S5}) hold if
$\hat q_\pm^{ii}(k_1,k_2,k_3)=f(k_1)f(k_2)f(k_3)$ with
$$
f(z)=
({(1-\cos(r_0 z/\sqrt 3))}/{z^2})^N, \,\,\,z\in\R,
$$
where $N\geq 0$ is an integer, $2N-s>1$ ($s=4-2i$).

\subsubsection{Non-Gaussian measures}
Let us choose some odd nonconstant functions
 $f^0,\,f^1\in C^4 (\R)$ with
bounded derivatives.
Let us define $\mu^*_0$ as the distribution of the random function
$
(f^0(Y^0(x)),  f^1(Y^1(x))),
$
where $(Y^0,Y^1)$ is a random function
with  the Gaussian distribution $\mu_0$ from the previous
example.
Then {\bf S0}, {\bf S1} and  {\bf S3} hold for $\mu^*_0$
with some appropriate functions $\nu_d$ since
corresponding mixing coefficients
$\phi^*_{d_1,d_2}(r)= 0$ for $r\geq r_0$.
Therefore,  {\bf S0} implies
for  the corresponding correlation functions
$Q^*_0(x,y)=0$ for $|x-y|\geq r_0$, so {\bf S2}
also holds.
The measure $\mu_0^*$
is not Gaussian
since
 the functions $f^0$, $f^1$ are bounded and nonconstant.

\setcounter{equation}{0}
\section{Application to Gibbs measures}

We apply Theorem \ref{t1.2}
to the case when $\mu_{\pm}$ are the  Gibbs measures (\re{5})
corresponding to
different positive
temperatures $T_-\not= T_+$.

\subsection{Gibbs measures}
We will define the Gibbs measures $g_{\pm}$ as the
Gaussian measures with
the correlation functions (cf. (\re{5}))
\be\la{4}
q_{\pm}^{00}(x-y)= -T_{\pm}{\cal E}(x-y),~~
q_{\pm}^{11}(x-y)= T_{\pm}\delta(x-y),~~
q_{\pm}^{01}(x-y)= q_{\pm}^{10}(x-y)= 0,
\ee
where $x,y\in\R^3$.
The correlation functions $q^{ij}_{\pm}$
do not satisfy condition {\bf S2} because of
singularity at $x=y$.
The singularity means that the measures $g_\pm$ are not concentrated
in the space $\cal H$.
Let us introduce appropriate functional spaces
for measures $g_\pm$.
First, let us define the weighted Sobolev space
with any $s,\alpha\in \R$.
\begin{definition}
$H_{s,\alpha}(\R^3)$
is the Hilbert space of the distributions
$u\in S'(\R^3)$ with the finite norm
\be\la{ws}
\Vert u\Vert_{s,\alpha}\equiv
\Vert \langle x\rangle^{\alpha}\Lambda^s u
\Vert_{L_2(\R^3)}<\infty,
\,\,\,\,\Lambda ^s u\equiv F^{-1}\Bigl[
\langle k\rangle^s \hat u(k)\Bigr].
\ee
\end{definition}

Let us fix arbitrary $s,\alpha<-3/2$.
\begin{definition}
$ G_{s,\alpha}$ is the Hilbert space
$H_{s+1,\alpha}(\R^3)\oplus H_{s,\alpha}(\R^3)$,
with the norm
$$
\brr Y\brr_{s,\alpha}\equiv
\Vert u\Vert_{s+1,\alpha}+\Vert v\Vert_{s,\alpha}<\infty,~~
Y= (u,v).
$$
\end{definition}

Introduce   the Gaussian Borel probability  measures
$g_{\pm}^0(du)$, $g_{\pm}^1(dv)$  in spaces
$ H_{s+1,\al}(\R^3)$ and  $H_{s,\al}(\R^3)$,
respectively,
 with characteristic
functionals
$$
\ba{c}
\left.
\ba{rcl}
\hat g_{\pm}^0(\psi)&=&\ds \int \ds \exp\{i\langle
 u,\psi\rangle\}g_{\pm}^0(du)=
\ds \exp\{\frac{\langle\De^{-1}\psi,\psi\rangle}{2\beta_\pm}\}\\
~&&\\
\hat g_{\pm}^1(\psi)&=&\ds \int \ds \exp\{i\langle v,\psi\rangle\}g_{\pm}^1(dv)=
\ds \exp\{-\frac{\langle\psi,\psi\rangle}{2\beta_\pm}\}
\ea\right|\psi\in D.
\ea
$$
By the Minlos theorem, \cite{CFS},
the  Borel probability measures $g^0_{\pm}$, $g^1_{\pm}$
exist in the spaces $ H_{s+1,\al}(\R^3)$, $H_{s,\al}(\R^3)$,
 respectively,
 because {\it formally} (see Appendix B)
\be\la{Min}
\int \Vert u\Vert^2_{s+1,\alpha} g_{\pm}^0(du)<\infty,
~~\int\Vert v\Vert^2_{s,\alpha} g_{\pm}^1(dv)
<\infty,~~s,\alpha<-3/2.
\ee

Finally, we define the Gibbs measures
$g_{\pm}(dY)$ as the Borel probability measures
$g^0_{\pm}(du)\times g^1_{\pm}(dv)$
 in
$G_{s,\al}$.
 Let $g_0(dY)$  be the  Borel probability measure in
$G_{s,\al}$
that is  constructed as in
  the Example of previous section with
$\mu_{\pm}(dY)=g_{\pm}(dY)$.
It satisfies {\bf S0} and {\bf S1}
with  $q_{\pm}^{ij}$ from (\ref{4}).
However, $g_0$ does not satisfy {\bf S2}.
Therefore, Theorem
\ref{t1.2} cannot be applied directly to $\mu_0=g_0$.
 $G_{s,\al}\subset {\cal H}^s$
by the standard arguments of pseudodifferential
equations, \ci{H3}.
The next lemma follows by Fourier transform from the finite speed
of propagation for wave equation.
\begin{lemma}
The operators $U(t):Y_0\mapsto Y(t)$
allow  a continuous
extension ${\cal H}^{s}\mapsto{\cal H}^{s}$.
\end{lemma}

\subsection{Convergence to equilibrium}
Let $Y_0$ be  the random function
 with the distribution $g_0$, hence
$Y_0\in G_{s,\al}$ a.s.
Denote by $g_t$ the distribution of $U(t) Y_0$.
\begin{theorem}\la{t1.2'}
Let $s<-{5}/{2}$. Then
there exists
a Gaussian  Borel probability measure  $g_\infty$ in
 ${\cal H}^{s}$ such that
\be\la{1.8g}
 g_t \,
{\buildrel {\hspace{2mm}{\cal H}^{s }}\over
{- \hspace{-2mm} \rightharpoondown }}\,
   g_\infty,\,\,\,\,t\to \infty.
\ee
\end{theorem}
{\bf Proof}
Let us fix an $s<-{5}/{2}$ and introduce  the  random function
$Y_0^s:=\Lam^s Y_0$,  $Y_0^s\in G_{0,\al}$ a.s.
Let us denote  by $g_t^s$ the distribution of $U(t)(\Lam^s Y_0),\,\,t\in\R$.
Then $g_t^s=g_t\Lam^{-s}$, and $\ds g_t=g_t^s\Lam^s$ since
  $\Lam^s(U(t) Y_0)= U(t)(\Lam^s Y_0)$.
Let us denote by $Q_t^s(x,y)$ the (matrix) correlation function of
measure $g_t^s$.

Measure  $g_0^s$ obviously
satisfies
{\bf S0}.
The correlation function $Q_0^s(x,y)$ also
satisfies
 {\bf S1} with a suitable modification:
(\re{3}) holds up to $\ds\de(1+|x|+|y|)^{-N}$ with any
$\delta, N>0$ and with $a=a(\de)$. This follows from the convolution
representation $Q_0^s(x,y)=Q_0(x,y)*(\Lam_s(x)\Lam_s(y))$ since
$\Lam_s(x)\equiv F^{-1}\langle k\rangle^s$
 is a function 
$\in L^1_{\rm loc}(\R^3)$
with a rapid long-range decay.
{\bf S2} also holds for $g_0^s$
with the functions
$\nu_d(r)=C(1+r)^{-1-d}$
for a sufficiently large  $C=C(s, T_\pm)>0$.
It follows immediately for $s<0$
with sufficiently large $|s|$
from the same convolution representation.
For $s<-5/2$ it follows by  the pseudodifferential
operators techniques.

Then the conclusions of Lemmas \re{lcom}, \re{p2.1} hold for
the random function $Y_0^s$  and the correlation
 functions $Q_t^s(x,y)$ of the measures $g_t^s$.
The proofs are almost unchanged.
Hence,
the convergence (\ref{1.8})  holds for the Gaussian measures
$g_t^s$: $\forall\ve>0$
\be\la{thet}
g_t^s\toHe  g_\infty^s,\,\,\,
t\to\infty,
\ee
where $g_\infty^s$
is a Gaussian measure in ${\cal H}$.
Therefore,
$$
 g_t \,
{\buildrel {\hspace{2mm}{\cal H}^{s-\ve }}\over
{- \hspace{-2mm} \rightharpoondown }}\,
   g_\infty,\,\,\,\,t\to \infty,
$$
since
$\ds g_t=g_t^s\Lam^s$.
This implies Theorem \ref{t1.2'}.
\hfill$\Box$
\medskip

The limiting measure $g_{\infty}$ is Gaussian with  the
correlation matrix $Q_{\infty}=
 (Q_{\infty}^{ij}(x,y))_{i,j= 0,1},$
where
\beqn
Q_{\infty}^{00}(x,y)\!\equiv\!&\!
 q_{\infty}^{00}(x-y)=&\!\!\! \!
- \frac{1}{2}(T_++T_-){\cal E}(x-y),\la{Q00}\\
Q_{\infty}^{10}(x,y)\!=\! -Q_{\infty}^{01}(x,y)
\!\equiv\!&\! q_{\infty}^{10}(x-y)=&
\frac{1}{2}(T_+-T_-)P(x-y),\la{Q10}\\
Q_{\infty}^{11}(x,y)\!\equiv\!&\! q_{\infty}^{11}(x-y)=&
\frac{1}{2}(T_++T_-)\delta(x-y).\la{Q11}
\eeqn
The identities (\ref{Q00})--(\ref{Q11}) follow formally
from (\ref{4}) and from (\ref{q00})--(\ref{q11}).
For the proof we apply (\ref{q00})--(\ref{q11}) to
the initial measure  $g_0^{s}$.

\subsection{Limit energy current density}
Let $u(x,t)$ be the random solution to (\ref{1}) with
the initial measure $\mu_0$ satisfying {\bf S0}--{\bf S3}.
The mean energy current density is
$E j(x,t)=-E \dot u(x,t)\na u(x,t)$.
Therefore,
in the limit $t\to\infty$,
$$
E j(x,t)\to  \ov j_\infty
= \nabla q_{\infty}^{10}(0).
$$
Respectively,
in the case of the ``Gibbs'' initial measure $g_0$,
the expression (\ref{Q10}) for the limiting correlation function
implies {\it formally} that
$$
\ov j_\infty= \frac{T_+-T_-}2 \nabla P(0),
$$
where
$[\nabla P](z)=
-F^{-1}\Bigl[\ds\frac{k\sgn k_3}{|k|}\Bigr](z)$.
Hence, formally we have the ``ultraviolet diverging''
limit mean energy current density,
$$
\ov j_\infty=
-\frac{T_+-T_-}{2(2\pi)^3}
\int\limits_{\R^3} \frac{k \sgn k_3}{|k|} dk=
-\infty\cdot
(0,0,T_+-T_-).
$$
On the other hand, for the convolution  $U(t) (Y_0*\theta)$
the corresponding limiting
mean energy current density is finite,
$$
\ov j_\infty^{\theta}=- \frac{T_+-T_-}
{2(2\pi)^3}\int\limits_{\R^3}
|\hat\theta(k)|^2 \frac{k
\sgn k_3}{|k|} dk=-C_\theta\cdot(0,0,T_+-T_-),
$$
if $\theta(x)$ is axially symmetric with respect to $Ox_3$;
$C_\theta >0$ if $\theta(x)\not\equiv 0$.

\setcounter{equation}{0}
\section{Compactness of the measures family}

Proposition \re{l2.1}  can be deduced from the bound (\re{2.1})
below with the help of the
Prokhorov Theorem \cite[Lemma II.3.1]{VF} as in
\ci[Theorem XII.5.2]{VF}.
\begin{lemma}\la{lcom}
Let {\bf S0}--{\bf S2} hold.
Then  the following bounds hold
 \beqn
 \sup\limits_{t\geq 0}
E \Vert U(t)Y_0\Vert^2_R<\infty,\,\,\,R>0.        \la{2.1}
\eeqn
\end{lemma}
{\bf Proof}\,\,Assumption
{\bf S2} and Proposition \re{p1.1} iii) imply by the Fubini Theorem
the existence of the 
 correlation functions in (\ref{qd}),
where $Y^i(x,t)$ are the components of
$Y(x,t)=(Y^0(x,t),Y^1(x,t))$.
Therefore, Definition (\re{d1.1}) implies
\beqn
E \Vert  Y(\cdot,t)\Vert^2_R&=&
E \int\limits_{|x|<R}|Y^0(x,t)|^2dx+
E\int\limits_{|x|<R}|\na Y^0(x,t)|^2dx+
E \int\limits_{|x|<R}|Y^1(x,t)|^2dx\nonumber\\
~\nonumber\\
&=&\!\int\limits_{|x|<R}Q_t^{00}(x,x)dx+
\!\int\limits_{|x|<R}\na_x\!\cdot\!\na_y Q_t^{00}(x,y)|_{y=x}dx+
\!\int\limits_{|x|<R}Q_t^{11}(x,x)dx.~~~\la{E0Q}
\eeqn
We bound  for example the integral  of $Q_t^{00}(x,x)$ in
(\ref{E0Q}) in the particular case when $Y^0_0\equiv u_0(x)=0$ almost surely.
The general  case will be considered in Section 6 as well as
the bounds for two remaining integrals in (\ref{E0Q}).
Let us assume for a moment that the function $Y^1_0\equiv v_0$
is continuous almost surely. Then the
Kirchhoff formula (\re{Kir}) gives by  the Fubini Theorem,
\be\la{r2.1}
Q_t^{00}(x,x)=
\frac{1}{(4\pi t)^2}\int\limits _{S_t(x)\times S_t(x)}
Q_0^{11}(x',x'')  dS(x') dS(x'').
\ee
Let us assume for a moment that
\be\la{H3'}
Q^{ij}_0(x',x'')= 0~~\mbox{for }~~|x'-x''|\ge r_0,~~ i,j= 0,1.
\ee
Then  (\ref{r2.1}) implies the uniform bound
\be\la{r2.1'}
Q_t^{00}(x,x)\le\frac{C}{t^2}
 \int\limits_{\ba{c}S_t(x)\times S_t(x)\\| x'-x''| \leq
r_0\ea}dS(x') dS(x'')\leq  I= C_1r_0^2,\,\,\,t\in\R.
\ee
Hence, the bound follows,
\be\la{r2.1''}
\int_{|x|<R}Q_t^{00}(x,x)dx\le CIR^3,~~t\in \R.
\ee
Next we remove the additional assumption (\ref{H3'})
by the following known lemma on spherical integral
identity, \ci{I}.
\begin{lemma}\la{linb}
Let $h(r)\in C(0,+\infty)$. Then
 for any $r_0\ge 0$ and $x''\in S_t(x)$ the identity holds,
\be\la{inb}
\int\limits_{\{x'\in S_t(x):\,|x'-x''|\ge r_0\}}
h (|x'-x''|)dS(x')=2\pi\int\limits_{r_0}^{2t} rh (r)dr.
\ee
\end{lemma}
Therefore, (\ref{r2.1}),
{\bf S2} with $d=2$ and Lemma \ref{linb} with
$r_0=0$   imply (see (\ref{phi})),
\be\la{r2.1'''}
Q_t^{00}(x,x)
\leq
\frac{C}{(4\pi t)^2}\int\limits _{S_t(x)\times S_t(x)}
\!\!\!\!\nu_2(|x'-x''|)\,dS(x')dS(x'')
\le C_1\int\limits_0^{2t}
r\,\nu_2 (r)\,dr  \le C_2 <\infty.
\ee
Then (\re{r2.1''}) follows without the assumption  (\ref{H3'}).
The assumption on the a.s. continuity of $v_0(x)$
can be removed by a convolution with a function
$\theta\in D$.
\hfill$\Box$

\setcounter{equation}{0}
\section{Convergence of correlation functions}
Here we prove the convergence (\ref{corf}) of the
correlation functions of measure $\mu_t$.
This implies  the convergence of
the characteristic functionals $\hat\mu_t$
in the case of Gaussian measures $\mu_0$, $\mu_{\pm}$.

\begin{lemma}\la{p2.1}
Let {\bf S0}--{\bf S2} hold.
 The following convergence holds as $t\to\infty$
\be\la{4.0}
Q^{ij}_{t}(x,y)\to
Q^{ij}_{\infty}(x,y),\,\,\,\forall x,y\in\R^3,\,\,\,\forall i,j= 0,1.
\ee
\end{lemma}
{\bf Proof}
We prove the lemma again for $i= j= 0$  in the particular case,
  $u_0\equiv 0$
almost surely.
The general case is considered in Section 6.
 Let us assume for a moment that the function $v_0(z)$
is continuous almost surely. Then the
Kirchhoff formula (\re{Kir}) and the Fubini Theorem give
\be\la{2.2}
Q_t^{00}(x,y)=
Eu(x,t)u(y,t)=
\frac{1}{(4\pi t)^2}\int\limits_{S_t(x)} dS(x')
\int\limits_{S_t(y)}   Q^{11}_0(x',y') dS(y').
\ee
This integral is the  convolution of  $Q^{11}_0(x,y)$
in both variables $x,y$ with a distribution
of compact support.
The convolution of distributions with compact
support is commutative.
Therefore, the assumption on the a.s. continuity of $v_0(x)$
can be removed by a convolution with a function
$\theta\in D$.
 Changing the variables $x'=x+\omega t$ in the right hand side
of (\ref{2.2}), we get
\beqn\la{2.2'}
&&\frac{1}{(4\pi t)^2}\int\limits_{S_t(x)}dS(x')
\int\limits_{S_t(y)}
Q^{11}_0(x',y') dS(y')\nonumber\\~&&\nonumber\\
&=&
\frac{1}{(4\pi )^2}
\int\limits_{|\omega|= 1,\omega_3<0}\!\!\!\!\!\!\!\! dS(\omega)
\int\limits_{S_t(y)}\!\!\!
Q_0^{11}(x+\omega t,y') dS(y')
+\frac{1}{(4\pi )^2}\int\limits_{|\omega|= 1,\omega_3>0}
\!\!\!\!\!\!\!\! dS(\omega)
\int\limits_{S_t(y)}\!\!\!
Q_0^{11}(x+\omega t,y') dS(y')\nonumber\\~&&\nonumber\\
&=& I_-(t,x,y)+I_+(t,x,y).
\eeqn
Let us recall that $\nu(r)\equiv\nu_2(r)$.
\begin{definition} \la{Cnu}
$C_\nu(\R^3)$ is the space of functions $
f(y)\in C(\R^3)$ s.t. $|f(y)|\le C\nu (|y|)$
with a constant $C\in\R$.
\end{definition}
Let us define for $f(y)\in C_\nu(\R^3)$
\be\la{Rf}
{\cal R}f(v)\equiv
\frac{1}{(4\pi )^2}
\int\limits_{|\omega|= 1,\pm\omega_3>0}\!\!\! dS(\omega)
\int\limits_{p\cdot\omega=v\cdot\omega}\!\!
f(p) d^2p,\,\,\,v\in\R^3.
\ee
Here $d^2p$ is the Lebesgue measure on the plane
 $p\cdot\omega= v\cdot\omega$.
Note that  the integrals with $\pm$ are identical and converge due to
(\re{phi}). Hence, the operator ${\cal R}:\, C_\nu(\R^3)\to C_b(\R^3)$
is continuous with the obvious norm in $C_\nu$:
$\Vert f\Vert_{C_{\nu}}=
\sup\limits_{y\in\R^3}\ds\frac{|f(y)|}{\nu(|y|)}$.

The convergence (\ref{4.0}) follows for
$i= j= 0$
from (\ref{2.2}), (\ref{2.2'}) and
Lemmas \ref{l3.2} and \ref{l3.3}.
\begin{lemma}\la{l3.2}
Let {\bf S2} hold. Then for $x,y\in\R^3$,
\be\la{2.20}
I_{\pm}(t,x,y)\to {\cal R}q_\pm^{11}(x-y),~~t\to\infty.
\ee
\end{lemma}
\begin{lemma}\la{l3.3}
 Let $f(y)\in C_\nu(\R^3)$. Then
\be\la{l3.31}
{\cal R} f=
-\frac{1}{4}{\cal E}*f.
\ee
\end{lemma}
 Lemma \ref{l3.3} is proved in Appendix~B.\\
~\\
{\bf Proof of Lemma \ref{l3.2}.}
For a moment
we assume  additionally
(\ref{H3'}).
Denote by $I_{11}$  the inner integral entering (\ref{2.2'}):
\be\la{inni}
I_{11}\equiv I_{11}(x,y,\omega,t)=
\int\limits_{S_t(y)}
Q_0^{11}(x+\omega t,y')~ dS(y').
\ee
Change  the variables  $y'= y+\omega t+p$ and denote $R=|x-y|$.
(\ref{H3'}) implies that
$Q_0^{11}(x+\omega t,y+\omega t+p)= 0$  for $|p|\ge r_0+R$,
hence (\ref{inni}) becomes
\be\la{innib}
I_{11}=
\int\limits_{S_t(-\om t)\cap  B_0}
Q_0^{11}(x+\omega t,y+\omega t+p)~ dS(p),
\ee
where $B_0$ denotes the ball $|p|\le r_0+R$.
The sphere $S_t(-\om t)$ contains the point $0$, hence
in a neighborhood of the origin
the sphere converges to its tangent plane $\om^\bot$
as $t\to\infty$.

Further, consider the case $\omega_3<0$ and $\omega_3>0$ separately.
For $\omega_3<0$   and  sufficiently
large  $t>t(\om)>0$,
$$
x_3+\omega_3 t<-a,~~~~ y_3+\omega_3 t +p_3<-a,
\,\,\,\,\,\,\mbox{  for  }\,\,\,|p|\le r_0+R.
$$
Then  {\bf S1}  implies that
\be\la{s2}
Q_0^{11}(x+\omega t,y+\omega t+p)=q_-^{11}(x-y-p)
\ee
Therefore, if $\omega_3<0$,
\be\la{innic}
I_{11}\to
\int\limits_{\om^\bot \cap  B_0}
q_-^{11}(x-y-p)~ d^2p, \,\,\,t\to\infty
\ee
that coincides with the inner integral in the right hand
 side of (\ref{Rf}), with $f=q_-^{11}$
and $v=x-y$. Similarly for $\omega_3>0$.
 Lemma \ref{l3.2} is proved
with the additional assumption (\re{H3'}).
At last, Lemma \re{linb} and {\bf S2} give the  uniform smallness of integral
(\re{inni}) over
$|p|\geq r_0+R$ with large $r_0$.
Therefore,  (\re{innic})
holds for any $\om$ with $\om_3\ne 0$.
Hence, (\re{2.20}) follows
by the Lebesgue Theorem on dominated convergence.
\hfill$\Box$

\setcounter{equation}{0}
\section{Correlation functions
in general case}
We prove Lemmas \re{lcom} and  \re{p2.1}
in the general case.
Let us assume for a moment that $u_0\in C^1(\R^3)$
and $v_0\in C(\R^3)$ almost surely. Then we
apply the general
Kirchhoff formula for the solution $u(x,t)$ to the Cauchy problem
(\ref{1}): formally,
\be\la{2.01}
u(x,t)= \frac{1}{4\pi t}\int\limits_{S_t(x)}
\Bigl(
v_0(x')+\frac{1}{t}u_0(x')+\nabla u_0(x')\cdot n_x(x')\Bigr)
  dS(x'),
\ee
where $n_x(x')=\ds\fr{x'-x}{|x'-x|}$.
It
implies  similarly to (\ref{2.2}),
\beqn
\hspace{-12mm}Q_t^{00}(x,y)=
\frac{1}{(4\pi t)^2}\int\limits_{S_t(x)} dS(x')
\int\limits_{S_t(y)}
\left(\Bigl[Q^{11}_0(x',y')+
\nabla_{y'}(\nabla_{x'} Q^{00}_0(x',y')\cdot n_x(x'))
\cdot n_y(y')\Bigr]\right.\nonumber\\
~\nonumber\\
\hspace{-10mm}+\frac{1}{t}\left[
Q^{10}_0(x',y')+
Q^{01}_0(x',y')+\nabla_{y'} Q^{00}_0(x',y')\cdot
n_y(y')+\nabla_{x'} Q^{00}_0(x',y')\cdot
n_x(x')+\frac{1}{t}Q^{00}_0(x',y')\right]\nonumber\\
~\nonumber\\
\hspace{-20mm}\left.+\Bigl[\nabla_{x'} Q^{01}_0(x',y')\cdot n_x(x')+
\nabla_{y'} Q^{10}_0(x',y')\cdot n_y(y')
\Bigr]\right) dS(y').\,\,\,\,\,\la{Qt00}
\eeqn
{\bf Proof of Lemma \re{lcom} in the general case}
We will prove the uniform bounds for $Q_t^{00}(x,x)$,
$\na_x\cdot\na_y Q_t^{00}(x,y)|_{x=y}$ and $Q_t^{11}(x,x)$. Then
(\re{E0Q}) implies (\re{2.1}).

{\it Step 1}~
(\re{Qt00}) represents
$Q_t^{00}(x,y)$ as the  sum of convolutions with
$D^{\alpha,\beta}_{x,y}Q^{kl}_0(x,y)$
in both variables $x,y$,
with
 $0\le d\equiv k+l+|\al|+|\beta|\le 2$.
Therefore,
$\na_x\cdot\na_y Q_t^{00}(x,y)$
is the similar sum involving
$D^{\alpha,\beta}_{x,y}Q^{kl}_0(x,y)$ with
 $2\le d\le 4$.
The similar representation holds for $Q_t^{11}(x,y)$.
Hence,
$\na_x\!\cdot\!\na_y Q_t^{00}(x,y)$ and
$Q_t^{11}(x,y)$ can be estimated
 by the method of the proof of Lemma \re{lcom}
in Section 4.
Indeed, due to  {\bf S2} with $d=2$,
 (\ref{phi}) and Lemma \re{linb}
 we get (cf. formula (\ref{r2.1'''})),
$$
\nabla_x\cdot\nabla_y  Q_t^{00}(x,y)|_{x=y}+
Q_t^{11}(x,x)
\leq
\frac{C}{(4\pi t)^2}\int\limits _{S_t(x)\times S_t(x)}
\!\!\!\!\nu_2(|x'-x''|)\,dS(x')dS(x'')
\le C_1 <\infty.
$$
{\it Step 2}~
$Q_t^{00}(x,y)$ requires the  particular attention
due to the presence in the integrand of the functions
  $D^{\alpha,\beta}_{x',y'}Q^{kl}_0(x',y')$
 that are estimated
by $\nu_d(|x'-y'|)$ with $d=0,1$.
In this case due to (\ref{phi})
and  Lemma \ref{linb}
we have to analyze  (\ref{Qt00}) more carefully.
The corresponding contribution
of  $D^{\alpha,\beta}_{x',y'}Q^{kl}_0(x',y')$
with $d=k+l+|\al|+|\beta|=0,1$
is
$$
I^{00}_t(x,y)
=
\ds\frac{1}{(4\pi t)^2}\int\limits_{S_t(x)}dS(x')
\int\limits_{S_t(y)}
\frac{1}{t}\left[Q^{10}_0(x',y')+
\dots+\frac{1}{t}Q^{00}_0(x',y')\right]dS(y').
$$
\begin{lemma}\la{lgibs}
The integral $I^{00}_t(x,y)$ converges to zero as $t\to\infty$.
\end{lemma}
{\bf Proof} The assumption {\bf S2} implies
\be\la{bgibs}
|I^{00}_t(x,y)|\le
\frac{C}{(4\pi t)^2}\int\limits_{S_t(x)}dS(x')
\int\limits_{S_t(y)}
\frac{1}{t}\left[
4\nu_1 (|x'-y'|)+\frac 1t\nu_0 (|x'-y'|)\right]dS(y').
\ee
Therefore,  Lemma \re{linb} implies
\beqn
|I^{00}_t(x,y)|&\le&
\frac{C}{(4\pi t)^2}\int\limits_{S_t(x)} dS(x')
\frac{1}{t}\left[\int_0^{2t}
(4r\nu_1 (r)+\frac 1t r\nu_0 (r))dr\right]\nonumber\\
&&~\nonumber\\
&\le&
C_1
\int_0^{2t}
(\frac rt\nu_1(r)+\frac r{t^2} \nu_0(r))dr.\la{ibgibs}
\eeqn
Now (\re{phi}) implies the convergence to zero
by the Lebesgue theorem.
\hfill$\Box$

Lemma \re{lcom} is proved
in the general case.\hfill$\Box$
\medskip\\
{\bf Proof of Lemma \re{p2.1} in the general case}
We will consider $i=j=0$. The other cases can be considered similarly.

{\it Step 1}~
The integrals of
$D^{\alpha,\beta}_{x',y'}Q^{kl}_0(x',y')$
with
$d\equiv k+l+|\alpha|+|\beta|\le 1$
entering
(\re{Qt00}),
converge to zero by Lemma \re{lgibs}.
For the integrals of $D^{\alpha,\beta}_{x',y'}Q^{kl}_0(x',y')$
with
$2\le d\le 4$,
the convergence
follows by the method of the proof of Lemma \re{l3.2}.
Let us define for
 the functions
$f\in C_\nu^1(\R^3):=
\{f\in L^1_{\rm loc}(\R^3): |\na f(y)|\in C_\nu(\R^3)\}$
(cf. Definition \ref{Cnu}),
the operator
\be
{\cal P}f(v):= \frac{1}{(4\pi)^2}
\int\limits_{|\omega|= 1,\omega_3>0}\!\!\!\! dS(\omega)
\int\limits_{v\cdot\omega= p\cdot\omega}\!\!
\na f(p)\!\cdot\!\om\,\, d^2p,\,\,\,v\in\R^3.\la{P}
\ee
With the obvious norm in $C_\nu^1$:
$\Vert f\Vert_{C_\nu^1}=
\sup\limits_{y\in\R^3}\ds\frac{|\nabla f(y)|}{\nu(|y|)}$
, the operator
${\cal P}:~C_\nu^1(\R^3)\to C_b(\R^3)$ is continuous.
For instance, the operator ${\cal P}$ can be applied to
$q^{kl}_{\pm}$ with $1\le k+l\le 2$
since $q^{kl}_{\pm}\in C^1_\nu(\R^3)$
by {\bf S2}.
Similarly,
the operator ${\cal R}$ (see formula (\ref{Rf}))
can be applied to
$D^{\al}q^{kl}_{\pm}$ with $2\le k+l+|\al|\le 4$
since  $D^{\al}q^{kl}_{\pm}\in C_\nu(\R^3)$.
Now, (\re{Qt00}) and
 the method of proof of Lemma \re{l3.2}  imply
 the convergence
(\ref{4.0}) with  $i=j=0$
to the limiting function
\be
\ q^{00}_*={\cal R}\Bigl[q_{+}^{11}+q_{-}^{11}-
\De( q_{+}^{00}+q_{-}^{00})\Bigr]+
{\cal P}\Bigl[ q_{+}^{01}-q_{-}^{01}-
 q_{+}^{10}+ q_{-}^{10}\Bigr].\la{q00a}
\ee
{\it Step 2}~
It remains to prove that $q^{00}_*=q^{00}_{\infty}$.
First, let us  prove that
\be\la{RD}
{\cal R}\De q_+^{00}=-\fr 14 q_+^{00}.
\ee
In fact, $\De ({\cal R}\De q_+^{00})=-\ds\fr 14\De q_+^{00}$
due to  (\ref{l3.31}), hence
$f(x)\equiv {\cal R}\De q_+^{00}-q_+^{00}$
is a smooth harmonic function in $\R^3$. On the other hand,
$\De q_+^{00}\in C_\nu(\R^3)$ by {\bf S2}. Hence,
 $g(x)\equiv {\cal R}\De q_+^{00}\in C_b(\R^3)$, and moreover,
\be\la{dec}
g(x)\to 0,\,\,\,|x|\to\infty.
\ee
Indeed,
\be\la{bau}
|\int\limits_{p\cdot\om=x\cdot\om}\De q_+^{00}(p)
d^2p|\le
\int\limits_{p\cdot\om=x\cdot\om}\nu (|p|)d^2p=
2\pi\int\limits_{x\cdot\om}^\infty r\nu (r)dr
\ee
similar to (\re{inb}) with $t=\infty$. This integral is bounded
uniformly in $|\om|=1$ and converges to zero if
$|x|\to\infty$ and
$x=|x|\theta$ with $\theta\cdot\om\not= 0$.
Therefore, (\re{dec}) follows from  (\re{Rf})
by the Lebesgue theorem.
Further, $|f(x)|\leq |g(x)|+\nu_0(|x|)$
again by {\bf S2}. At last, $\nu_0(r_n)\to 0$ for some sequence
$r_n\to\infty$ due to (\re{phi}).
Finally, the maximum principle and (\re{dec}) imply
for any fixed $x\in\R^3$,
$$
|f(x)|\le \max\limits_{|y|=r_n}|g(y)|+\nu_0(r_n)\to 0,\,\,\,
n\to\infty.
$$
Therefore, $f(x)\equiv 0$ and (\re{RD})
is proved.
Further, let us consider the terms with ${\cal P}$ in
 (\ref{q00a}).
Obviously, ${\cal P}f$ is a convolution.
We prove the next lemma in Appendix A.
Let us recall that
$P(x)=\ds-i
F^{-1}\Bigl[\ds\frac{\sgn k_3}{|k|}
\Bigr]$.
\begin{lemma}\la{l3.3'}
For $f\in D$ we have
\be\la{l3.32}
{\cal P}f=
\frac{1}{4}P *f.
\ee
\end{lemma}

Let us assume for a moment that
all the correlation functions $q^{kl}_\pm(\cdot)$
are smooth and have a rapid decay.
Then (\ref{q00a})
coincides with (\ref{q00})
 by (\ref{RD}) and
Lemmas \ref{l3.3}, \ref{l3.3'}.
In the  general case we consider  the formula
(\ref{l3.32}) as the definition
of the convolutions with $P$,
 entering (\ref{q00})--(\ref{q11}).
Lemma \re{p2.1} is proved
in the general case.\hfill$\Box$

\setcounter{equation}{0}
\section{Bernstein's argument for the wave equation}
In this and the subsequent section we develop
a version of the 
Bernstein `room-corridor'
method.
We use the standard integral
representation  for the solutions,
 divide the domain of integration
into `rooms' and `corridors' and evaluate
their contribution. As the  result, $\langle  U(t)Y_0,\Psi\rangle $
is represented
as the sum of
weakly dependent random variables.
We  evaluate  the variances  of these  random variables
that will be important in  next section.

For the wave equation the  similar method  has been used in
\ci[Section 6]{DKR}
for an odd $n\ge 3$.
Our  mixing condition {\bf S3}
is different from  \ci{DKR}
 (see Remark \re{rQ}).
Respectively, the method of \ci{DKR}
requires a suitable modification.

 Denote by ${\cal E}_t(x)\equiv {\cal E}(x,t)
=\ds\frac{1}{2\pi} \delta(|x|^2-t^2)$
the fundamental solution to the wave equation.
The support of ${\cal E}_t$ is the sphere
$S_t=\{x\in\R^3: |x|=t\}$. Therefore,
the dynamical group $U(t)$ of the problem (\ref{1'})
 is the convolution operator
\be\la{z.4'}
U(t) Y_0= {\cal G}_t*Y_0,\,\,t>0,
\ee
where
\be\la{z.41'}
{\cal G}_t=\left(
 \ba{lr}
 \dot {\cal E}_t & {\cal E}_t   \\
 \triangle {\cal E}_t & \dot {\cal E}_t
 \ea     \right).
\ee
Next we introduce a `room-corridor'  partition of the space $\R^3$.
Given $t>0$,
choose
$d\equiv d_t\ge 1$ and
$\rho\equiv\rho_t>0$
and an integer $N\equiv N_t>0$.
Asymptotic relations between $t$, $d_t$ and  $\rho_t$
are specified below.
Define
\be\la{rom}
a_1=-t,~b_1=a_1+d;\,\,\,
a_2=b_1+\rho,~b_2=a_2+d;\,\,\,\dots,~b_N\equiv a_N+d=t.
\ee
We divide  the sphere $S_t$  by the planes
orthogonal to the axis $Ox_3$  into the slabs
which we call the "rooms" $R_k^t$ $(k=1,...,N)$,
separated by the
"corridors" $C_k^t$ $(k=1,...,N-1)$,
\be\la{cr}
R^t_k=\{x\in S_t:~x_3\in[a_k,b_k]\},\,\,\,\,
C^t_k=\{x\in S_t:~x_3\in[b_k,a_{k+1}]\}.
\ee
Here $x=(x_1,x_2,x_3)$, $d$ is the width
of a room, and $\rho$ of a corridor.
Then
\be\la{un}
S_t=\Big( \cup R^k_t  \Big)\cup\Big( \cup C^k_t \Big).
\ee
For  any region $\Sigma\subset S_t$
we define the distribution
${\cal E}_{t,\Sigma} $
with the support in $\Sigma$
$$
\langle{\cal E}_{t,\Sigma}, \theta\rangle:=
\frac{1}{4\pi t}\int\limits_{\Sigma} \theta (z)dS(z),
\,\,\,\,\theta\in D.
$$
Note that $\forall t>0$
$
\dot {\cal E}_t(x)=
-\ds\fr{t}{\pi}\delta'(|z|^2-t^2)=
\fr{1}{t}{\cal E}_t(x)-
\nabla(\fr{x}{t} {\cal E}_t(x)).$
For  any region $\Sigma\subset S_t$
we define the distribution
$\dot{\cal E}_{t,\Sigma}$:
\be\la{dotE}
\dot {\cal E}_{t,\Sigma}(x):=
\fr{1}{t}{\cal E}_{t,\Sigma}(x)-
\nabla(\fr{x}{t}{\cal E}_{t,\Sigma}(x)).
\ee
Then for $\Sigma=S_t$ we have
$\dot {\cal E}_{t,\Sigma}=\dot {\cal E}_{t}$.
Let us denote
\be\la{z.4''}
 {\cal G}_{t,\Sigma}:=
\left(
 \ba{lr}
 \dot {\cal E}_{t,\Sigma}& {\cal E}_{t,\Sigma}   \\
\triangle {\cal E}_{t,\Sigma} & \dot {\cal E}_{t,\Sigma}
 \ea     \right).
\ee
We define the random variable
\be\la{Isigma}
I_t(\Si)=\langle{\cal G}_{t,\Si}*Y_0,\Psi\rangle,
\ee
where $\Psi\in {\cal D}$
is a fixed function from (\re{2.6}).
For instance, define
\be\la{Irc}
r_t^k=I_t(R_t^k),\,\,\,
c_t^k=I_t(C_t^k).
\ee
 (\re{un}) implies that
\be\la{razl}
\langle U(t)Y_0,\Psi\rangle= \langle{\cal G}_t*Y_0,\Psi\rangle=
\sum\limits_{k=1}^{N_t}r_t^k+
\sum\limits_{k=1}^{N_t-1}c_t^k.
\ee
\begin{lemma}  \la{l50.1}
    Let  {\bf S0}, {\bf S3} hold.
The following bounds hold for $t>1$ and $\forall k$
\beqn
 E|r^k_t|^2\le C(\Psi)\,d_t/t,\la{Er}
\\
 E|c^k_t|^2\le C(\Psi)~\rho_t/t.\la{Ec}
\eeqn
\end{lemma}
{\bf Proof}.
We prove the following estimate: for any
region $\Sigma\subset S_t$
\be\la{I_tSigma}
E|I_t(\Sigma)|^2\le C(\Psi)|\Sigma|/t^2.
\ee
Then  (\ref{Er}) and (\ref{Ec})
would follow from  this estimate with  $\Sigma=R_t^k$
and $\Sigma=R_t^k$, respectively,
 as
$|R_t^k|=2\pi t d_t$ and $|C_t^k| = 2\pi t\rho_t$.

Now we prove (\ref{I_tSigma}).
From
(\ref{z.4''}) and  (\ref{Isigma})
it follows
 that for $\Psi=(\Psi^0,\Psi^1)\in{\cal D}$
\be\la{111}
I(\Sigma)=
\langle \dot{\cal E}_{t,\Sigma}* u_0,\Psi^0
\rangle  +
\langle {\cal E}_{t,\Sigma}* v_0,\Psi^0\rangle
-
\langle {\cal E}_{t,\Sigma}*\nabla u_0,\nabla\Psi^1
\rangle  +
\langle \dot{\cal E}_{t,\Sigma}* v_0,\Psi^1\rangle.
\ee
Substituting (\ref{dotE}) in the first and the last terms
in the RHS of (\ref{111}), we get
\beqn
I(\Sigma)&=&
\langle {\cal E}_{t,\Sigma}* u_0/t,\Psi^0
\rangle  -
\langle \Big(\frac{x}{t} {\cal E}_{t,\Sigma}\Big)
*\nabla u_0,
\Psi^0\rangle  -
\langle {\cal E}_{t,\Sigma}*\nabla u_0,\nabla\Psi^1\rangle\nonumber\\
&&+\langle {\cal E}_{t,\Sigma}* v_0,\Psi^0\rangle
+
\frac{1}{t}
\langle {\cal E}_{t,\Sigma}* v_0,\Psi^1\rangle
+
\langle \Big(\frac{x}{t} {\cal E}_{t,\Sigma}
\Big)* v_0,
\nabla\Psi^1
\rangle.\nonumber
\eeqn
Hence,
\be\la{I(Sigma)}
I_t(\Sigma)
=\sum\limits_{j=1}^{M}
 I_t^{j},\,\,\, \mbox{where}\,\,\,
I_t^{j}=c_j(t)
\langle\bar {\cal E}^j_{t,\Sigma}*w_j,\theta_j\rangle.
\ee
Here $M\le 6$,
$c_j(t)$ is a bounded function for $t\ge\delta>0$,
$w_j$ is one of
 $t^{-1}u_0$, $\nabla u_0$ or $v_0$,
 $\bar {\cal E}^j_{t,\Sigma}$
 is one of
${\cal E}_{t,\Sigma}(z)$
or $\ds\frac{z}{t}{\cal E}_{t,\Sigma}(z)$,
$\theta_j$ is one of
$D^\alpha\Psi^l$ with $|\alpha|\le 1$.
Therefore,  for $t>1$
\beqn\la{Isigma1}
\!\!\!\!\!\!\!\!\!E |I_t(\Sigma)|^2 &\le&
C\sum\limits_{j=1}^{M}
E|I_t^j|^2\le
C\sum\limits_{j=1}^{M}
\langle E\Big[\Big(\bar {\cal E}^j_{t,\Sigma}*w_j\Big)(x)\,
\Big(\bar {\cal E}^j_{t,\Sigma}*w_j\Big)(y)\Big],
\theta_j(x)\theta_j(y)\rangle\nonumber\\
&\le&
C_1\sum\limits_{j=1}^{M}
\frac{1}{ t^2}
|\langle\int\limits_{\Sigma}
\int\limits_{\Sigma}
\Bigl(
\frac{1}{t^2}Q_{0}^{00}(x\!-z-\!(y-\!p))+
\!\!\!\!\!\!\sum\limits_{|\al|=|\beta|=1}
\!\!\!\!D^{\al,\beta}_{z,p}Q^{00}_0(x\!-\!z-\!(y-\!p))\\
&&
+Q^{11}_0(x\!-\!z-\!(y-\!p))
\Bigr)
\,dS(z)dS(p),\theta_j(x)\theta_j(y)\rangle|.\nonumber
\eeqn
We have
\be\la{r0}
\supp \Psi \subset B_{r_0}=
\{x\in\R^3:~|x|\le r_0\}
\ee
with an $r_0>0$.
Since
$x,y\in\supp\theta_j\subset\supp\Psi\subset B_{r_0}$,
$|x-z-y+p|\ge(|z-p|-2r_0)_+ $,
where $s_+=\max(s,0)$, $s\in\R$.
Since $\nu_d(r)$ are non-increasing functions,
(\ref{Isigma1})
and {\bf S3} imply
\beqn\la{sigma0}
\!\!\!\!\!\!\!\!\!\!E |I_t(\Sigma)|^2\!\!\! &\le&
\!\!\!\!C(\Psi)\frac{1}{ t^2}
\int\limits_{\Sigma}dS(z)
\int\limits_{\Sigma}
\Bigl(
\frac{1}{t^2}
\nu_{0}((|z-p|-2r_0)_+)+\nu_2((|z-p|-2r_0)_+)\Bigr)
\,dS(p).
\eeqn
Then
Lemma \ref{linb} and (\ref{phi}) imply
as in Lemma \ref{lgibs},
$$
\,\,\,\,\,\,\,\,\,\,\,\,\,\,\,\,\,\,\,\,\,\,
\,\,\,\,\,\,\,\,
E |I_t(\Sigma)|^2
\le
C\frac{|\Sigma|}{t^2}\int\limits_{0}^{2t}
\Bigl(
r\nu_2((r-2r_0)_+)+
\frac{r}{t^2}\nu_0((r-2r_0)_+) \Bigr)\,dr\le
 C_1\frac{|\Sigma|}{t^2}.
\,\,\,\,\,\,\,\,\,\,\,\,\,\,\,\,\,\,\,\,\Box
$$

\setcounter{equation}{0}
\section{Convergence of characteristic functionals}
In this section we complete the proof of Proposition \ref{l2.2}.
If ${\cal Q}_{\infty}(\Psi,\Psi)=0$
Proposition \ref{l2.2}  is obvious,
 due to (\ref{4.0}). Thus, we may assume that
\be\la{50.*}
{\cal Q}_{\infty}(\Psi,\Psi)\not=0.
\ee
Choose  $0<\delta<1$, and
\be\la{rN}
 N_t\sim (\ln (t+1))^{1/10},\,\,\,
\rho_t\sim t^{1-\delta},\,\,\,t\to\infty.
\ee
\begin{lemma}\la{r}
The following limit holds true:
\be\la{7.15'}
N_t\Bigl(
\nu_0^2(\rho_t)+\Bigl(
\frac{\rho_t}{t}\Bigr)^{1/2}\Bigr)
+
N_t^2\Bigl(
\nu_0(\rho_t)+\frac{\rho_t}{t}\Bigr)
\to 0 ,\quad t\to\infty.
\ee
\end{lemma}
{\bf Proof}.
Since  $\nu_d(r)$
are non-increasing functions,
 (\ref{phi}) implies
$$
\nu_0(r)\ln(r+1)=\int\limits_{0}^{r}
\frac{\nu_0(r)}{s+1}\,ds\le
\int\limits_{0}^{r}
\frac{\nu_0(s)}{s+1}\,ds\le C<\infty.
$$
Then (\ref{rN}) implies  (\ref{7.15'}).
\hfill$\Box$\\

By the triangle inequality,
\beqn
|\hat\mu_t(\Psi) -\hat\mu_{\infty}(\Psi)|&\le&
|E\exp\{i \langle U(t)Y_0,\Psi\rangle \}-
E\exp\{i{\sum}_t r_t^k\}|+\nonumber\\
&&+|\exp\{-\frac{1}{2}{\sum}_t E(r_t^k)^2\} -
\exp\{-\frac{1}{2} {\cal Q}_{\infty}(\Psi,\Psi)\}|+\nonumber\\
&&+ |E \exp\{i{\sum}_t r_t^k\} -
\exp\{-\frac{1}{2}{\sum}_t E(r_t^k)^2\} |\nonumber\\
&\equiv& I_1+I_2+I_3,\la{4.99}
\eeqn
where the sum ${\sum}_t$  stands for $\sum\limits_{k=1}^{N_t}$.
We are going to show that all the summands
 $I_1$, $I_2$, $I_3$ tend to zero as  $t\to\infty$.\\
{\it Step (i)}  Eqn (\ref{razl})   implies
\be\la{101}
 I_1=|E\exp\{i{\sum}_t r^k_t \}
\ds{(\exp\{i{\sum}_t c^k_t\}-1)|\le
 {\sum}_t E|c^k_t|\le
{\sum}_t
(E|c^k_t|^2)^{1/2}}.
\ee
>From (\ref{101}), (\ref{Ec})  and
(\ref{rN}) we obtain that
\be\la{103}
I_1\le C N_t(\rho_t/t)^{1/2}\to 0,~~t\to \infty.
\ee
{\it Step (ii)} By the triangle inequality,
\beqn
I_2&\le&
\frac{1}{2}\,|{\sum}_t E(r_t^k)^2-
 {\cal Q}_{\infty}(\Psi,\Psi)\!|\le
 \frac{1}{2}\,
|\!{\cal Q}_{t}(\Psi,\Psi)-{\cal Q}_{\infty}(\Psi,\Psi)\!|
\nonumber\\
&&+ \frac{1}{2}\, |E\Bigl({\sum}_t r_t^k\Bigr)^2
-{\sum}_t E(r_t^k)^2| +
 \frac{1}{2}\, |E\Bigl({\sum}_t r_t^k\Bigr)^2
-{\cal Q}_{t}(\Psi,\Psi)\!|\nonumber\\
&\equiv& I_{21} +I_{22}+I_{23}\la{104},
\eeqn
where ${\cal Q}_t$ is the  quadratic form with the integral kernel
$\Big(Q^{ij}_t(x,y)\Big)$.
Eqn (\ref{4.0}) implies $I_{21}\to 0,$ $t\to\infty$.
As to $I_{22}$, we first obtain that
\be\la{i22}
I_{22} \equiv
 \frac{1}{2}\, |E\Bigl({\sum}_t r_t^k\Bigr)^2
-{\sum}_t E(r_t^k)^2|
\le \sum\limits_{k< l}
| Er_t^k r_t^l|.
\ee
The next lemma is a corollary of (\ci[Lemma 17.2.3]{IL}).
\begin {lemma}\la{il}
 Let $\xi$ be a random value measurable with respect to the 
$\sigma$-algebra $\sigma_{d_1}({\cal A})$,
 $\eta$ be a random value  measurable with respect to the 
$\sigma$-algebra $\sigma_{d_2}({\cal B})$,
and dist$({\cal A}, {\cal B})\ge h>0$. \\
i) Let $(E|\xi|^2)^{1/2}\le a$, $(E|\eta|^2)^{1/2}\le b$. Then
$$
|E\xi\eta-E\xi E\eta|\le
C\, ab~
\phi^{1/2}_{d_1,d_2}(h).
$$
ii) Let $|\xi|\le a$, $|\eta|\le b$ almost surely. Then
$$
|E\xi\eta-E\xi E\eta|\le C\,ab~
\phi_{d_1,d_2}(h).
$$
\end{lemma}
\bigskip

We apply Lemma \ref{il} to deduce  that
$I_{22}\to 0$ as $t\to\infty$.
Note that
$r_t^k=$
$\langle{\cal G}_{t,R_t^k}*Y_0,\Psi \rangle$
 is measurable
with respect to the $\sigma$-algebra  $\sigma_{d_k}({\cal A}^k)$,
where
$$
{\cal A}^k=\{x-y:\,\,y\in
R_t^k,\,\,x\in\supp\Psi\subset B_{r_0}\}.
$$
The distance
between the different rooms $R_t^k$ is greater or equal to
$\rho_t$ according to
 (\ref{rom}) and (\ref{cr}). Then
 $\rho({\cal A}^k,{\cal A}^l)\ge\rho(R_t^k,R_t^l)-
2r_0\ge \rho_t-2r_0$.
Hence (\ref{i22}) and {\bf S0}, {\bf S3} imply, together with
Lemma \ref{il} i), 
\be\la{i222}
I_{22}\le
C N_t^2\nu_0((\rho_t-2r_0)_+)\to 0,~~t\to\infty,
\ee
because of (\ref{Er}) and
Lemma \ref{r}.
Finally, it remains to check  that $I_{23}\to 0$,
$t\to\infty$.
By
Cauchy-Schwartz inequality,
\beqn
I_{23}&\le&
 |E\Bigl({\sum}_t r_t^k\Bigr)^2
- E\Bigl({\sum}_t r_t^k +
{\sum}_t c_t^k\Bigr)^2 |\nonumber\\
& \le&
 N_t{\sum}_t E |c_t^k|^2  +
2\Bigl(
E({\sum}_t r_t^k)^2\Bigr)^{1/2}
\Bigl(
N_t{\sum}_t E|c_t^k|^2\Bigr)^{1/2}.\la{107}
\eeqn
 (\ref{Er}), (\ref{i22}) and (\ref{i222})
imply  $E({\sum}_t r_t^k)^2\le
 C_1+C_2N^2_t\nu_0((\rho_t-2r_0)_+)\le C_3<\infty$. Then
(\ref{Ec}),  (\ref{107}) and Lemma \ref{r} imply
\be\la{106'}
I_{23}\le C_1  N_t^2\rho_t/t+C_2 N_t(\rho_t/t)^{1/2} \to 0,~~t\to \infty.
\ee
So, $I_{21}$, $I_{22}$, $I_{23}$ tend to zero, as $t\to\infty$.
Then
 (\ref{104}) implies 
\be\la{108}
I_2\le
\frac{1}{2}\,
|{\sum}_t E(r_t^k)^2-
 {\cal Q}_{\infty}(\Psi,\Psi)\!|
\to 0,~~t\to\infty.
\ee
{\it Step (iii)}
It remains to verify
\be\la{110'}
I_3\equiv| E\exp\{i{\sum}_t r_t^k\}
-\exp\{-\fr12{\sum}_t E(r_t^k)^2\}| \to 0,~~t\to\infty.
\ee
Using Lemma \ref{il}, ii)
we obtain:
\beqn
&&|E\exp\{i{\sum}_t r_t^k\}-\prod\limits_{k=1}^{N_t}
E\exp\{i r_t^k\}|
\nonumber\\
&\le&
|E\exp\{ir_t^1\}\exp\{i\sum\limits_{k=2}^{N_t} r_t^k\}  -
 E\exp\{ir_t^1\}E\exp\{i\sum\limits_{k=2}^{N_t} r_t^k\} |
\nonumber\\
&&+
|E\exp\{ir_t^1\}E\exp\{i\sum\limits_{k=2}^{N_t} r_t^k\}
-\prod\limits_{k=1}^{N_t}
E\exp\{i r_t^k\}|
\nonumber\\
&\le& \nu^2_0((\rho_t-2r_0)_+)+
|E\exp\{i\sum\limits_{k=2}^{N_t} r_t^k\}
-\prod\limits_{k=2}^{N_t}
E\exp\{i r_t^k\}|.\nonumber
\eeqn
We then apply Lemma \ref{il}, ii) recursively and get,
according to  Lemma \ref{r},
\be\la{7.24'}
|E\exp\{i{\sum}_t r_t^k\}-\prod\limits_{k=1}^{N_t}
E\exp\{i r_t^k\}|
\le
 N_t\nu^2_0((\rho_t-2r_0)_+)\to 0,\quad t\to\infty.
\ee
It remains to verify the convergence
\be\la{110}
|\prod\limits_{k=1}^{N_t} E\exp\{i r_t^k\}
-\exp\{-\fr12{\sum}_t E(r_t^k)^2\}| \to 0,~~t\to\infty.
\ee
According to the standard statement of the
Central Limit Theorem (see, e.g. \ci[Thm 4.7]{P})
it suffices to verify the  Lindeberg condition:
$\forall\ve>0$
\be\la{Lind}
\frac{1}{\sigma_t}  {\sum}_t
 E_{\ve\sqrt{\sigma_t}}
|r_t^k|^2 \to 0,~~t\to\infty.
\ee
Here
$
\sigma_t\equiv{\sum}_t E( r_k^t)^2$,
and $E_\de f\equiv E(X_\de f)$,
where $X_\de$ is  the indicator of the event
$|f|>\de^2$.
Note that (\ref{108}) and (\re{50.*})
imply 
$$
\sigma_t \to
{\cal Q}_{\infty}(\Psi,\Psi)\not= 0,~~t\to\infty.
$$
Hence it remains to verify that $\forall\ve>0$
\be\la{zL}
{\sum}_t
E_{\ve}
|r_t^k|^2 \to 0,~~t\to\infty.
\ee
We check (\ref{zL}) in Sections 9, 10.
Finally,
(\ref{4.99}) and (\ref{103}), (\ref{108})-(\ref{110}) imply
 Proposition \ref{l2.2}.
\hfill$\Box$

\setcounter{equation}{0}
\section{The Lindeberg condition}
The proof of (\ref{zL}) can be reduced
to the case when
for some $b\ge 0$ we have, almost surely that
\be\la{boun}
|Y_0(x)|\le b,\quad\quad x\in\R^3.
\ee
The general case can be covered by  the 
standard cutoff argument in the following
way.
We decompose $Y_0$ in two summands:
the first one, satisfying
 the estimate (\ref{boun}), and
 the remainder.
For  large $b$,
 the dispersion of the remainder
is  small
due to {\bf S2}, {\bf S3}
and Lemma \ref{il}, i),
then the dispersion (\ref{Er}) of
the  corresponding variables $r_t^k$
is small uniformly in $t$.
The last  fact follows
from the proof of (\ref{Er}).

Further, we estimate
\be
{\sum}_t
E_{\ve}|r_t^k|^2=
{\sum}_t|R_t^k|~
\frac{1}{|R_t^k|}E_{\ve}|r_t^k|^2
\le 4\pi t^2
\max\limits_{k=1,..,N_t}
\frac{1}{|R_t^k|}E_{\ve}|r_t^k|^2.
\nonumber
\ee
Therefore, it remains to prove
\be\la{zll0}
\max\limits_{k=1,..,N_t}
\frac{1}{|R_t^k|}
E_{\ve}
|r_t^k|^2
 =o(t^{-2}),~~t\to\infty.
\ee
The Chebyshev inequality implies
\be\la{Cheb}
E_{\ve}|r_t^k|^2
\le \frac{1}{\ve^2}
E|r_t^k|^4.
\ee
Using (\ref{I(Sigma)}), we get
\be
E|r_t^k|^4=
E|I_t^1+...+I_t^M|^4
\le C(M)E(|I_t^1|^4+...+|I_t^M|^4).\la{112}
\ee
Therefore,  (\ref{zll0}) follows from the
   estimate
\be\la{zll}
\max\limits_{k}\frac{1}{|R_t^k|}
E|<\bar {\cal E}_{t,R_t^k}*w_k,\theta_k>|^4
=o(t^{-2}),\quad t\to\infty.
\ee
We prove the following proposition in the next section.
\begin{pro}\la{11.2}
Let (\re{boun}) holds, and
$w=t^{-1}u_0$,$\nabla u_0$ or $v_0$. Then
for any $\Sigma\subset S_t$
the bound holds
\be\la{zes}
E|\!\!<\bar {\cal E}_{t,\Sigma}*w,\theta>\!\!|^4
\le C(\theta)\left(\frac{b}{t}\right)^4 |\Sigma|^2.
\ee
Here $\bar {\cal E}^j_{t,\Sigma}$
 is one of
${\cal E}_{t,\Sigma}(z)$
or $\ds\frac{z}{t}{\cal E}_{t,\Sigma}(z)$;
$\theta_j$ is one of
$D^\alpha\Psi^l$ with $|\alpha|\le 1$.
\end{pro}
This proposition implies (\ref{zll}):
$$
\frac{1}{|R_t^k|}
E|<\bar {\cal E}_{t,R_t^k}*w_k,\theta_k>|^4
\le
 \frac{1}{|R_t^k|}
C(\Psi)\left(\frac{b}{t}\right)^4|R_t^k|^2
\le C(b,\Psi)\frac{|R_t^k|}{t^4}=o(t^{-2}),
$$
since $|R_t^k|\le 4\pi t^2/ N_t$,
 where $N_t\to\infty$. (\ref{zL}) is  proved.\hfill$\Box$

\setcounter{equation}{0}
\section{The fourth order moment functions}
We deduce
Proposition \re{11.2} from the bounds for
the fourth order moment functions.

Denote by
$m_{0}^{(l)}(\bar z):=Ew(z_1)\cdot...\cdot w(z_l)$,
$\bar z=(z_1,...,z_l)$, where
 $w(z_k)=v_0(z_k)$ for every $k=1,...,l$, or $w(z_k)=
\nabla u_0(z_k)$ for every  $k=1,...,l$,
or $w(z_k)=t^{-1}u_0(z_k)$
for every  $k=1,...,l$.
We have
 $\supp\theta\subset B_{r_0}$
for an $r_0>0$.
Then left hand side of  (\ref{zes}) is
estimated as follows,
\be\la{114}
E|<\bar {\cal E}_{t,\Sigma}*w,\theta>|^4\le
\frac{C(\theta)}{t^4}
\int\limits_{B_{r_0}^4}
\int\limits_{\Sigma^4}|m_{0}^{(4)}(\bar x-\bar z)|
dS(\bar z)\,d\bar x,
\ee
where $dS(\bar z):=dS(z_1)\dots dS(z_4)$.
Therefore, we have to prove that
\be\la{ix}
I(\ov x)\equiv\int\limits_{\Si^4} |m_0^{(4)}(\ov x-\ov z)|dS(\ov z)\le
Cb^4|\Si|^2,\,\,\,\ov x\in B_{r_0}.
\ee
{\it Step 1}
Let us
prove an estimate for the moment functions
$m_0^{(4)}(y_1,y_2,y_3,y_4)$
by the method
\ci[Section 6.2]{DKR}.
We use the mixing condition for
different configurations of the points
$y_1,y_2,y_3,y_4$
in the space $\R^3$.
\begin{lemma}
The bound holds
\beqn
\!\!\!\!\!\!&&|m_0^{(4)}(y_1,y_2,y_3,y_4)|\le
4b^4\left(
\nu^2_2(\frac{1}{3}|y_1-y_2|)+\nu^2_0(\frac{1}{3}
|y_1-y_2|)
t^{-4}\right)\la{m1} \\
&&+16b^4
\sum\limits_{i,j=0,2}\left(\frac{1}{t^i}\nu^2_{2-i}
(|y_1-y_3|)
\cdot \frac{1}{t^j}\nu^2_{2-j}(|y_2-y_4|)
+
\frac{1}{t^i}\nu^2_{2-i}(|y_1-y_4|)
\cdot \frac{1}{t^j}\nu^2_{2-j}(|y_2-y_3|)\right).
\nonumber
\eeqn
\end{lemma}
{\bf Proof}.
Let us
 divide the  space $\R^3$
in three regions $I_1, I_2, I_3$ by two
hyperplanes that are orthogonal to
the segment
$[y_1,y_2]$ and divide it in three equal segments,
$y_1\in I_1$, $y_2\in I_3$.
At least one of the regions $I_1, I_2, I_3$
 does not contain
$y_3, y_4$.
If the points $y_3, y_4\not\in I_1$, then {\bf S0}
 and {\bf S3} imply (\ref{m1}), since
\beqn
|m_0^{(4)}(y_1,y_2,y_3,y_4)|&=&
|m_0^{(4)}(y_1,y_2,y_3,y_4)-
m_0^{(1)}(y_1)m_0^{(3)}(y_2,y_3,y_4)|
\nonumber\\
&\le&
4b^4\left(
\nu^2_2(\frac{1}{3}|y_1-y_2|)+\nu^2_0(\frac{1}{3}|y_1-y_2|)
t^{-4}\right).\nonumber
\eeqn
The same proof is valid for the case
$y_3, y_4\not\in I_3$.
Now let us assume that
 $y_3, y_4\not\in I_2$,  for instance, $y_3\in I_1$,
$y_4\in I_3$. Then {\bf S0}, {\bf S3} imply
 (\ref{m1}), since
by Lemma \ref{il}, ii)
$$
|m_0^{(4)}(y_1,y_2,y_3,y_4)|\le
|m_0^{(4)}(y_1,y_2,y_3,y_4)-
m_0^{(2)}(y_1,y_3)
m_0^{(2)}(y_2,y_4)|
+
|m_0^{(2)}(y_1,y_3)
m_0^{(2)}(y_2,y_4)|
$$
$$
\le
4b^4\left(
\nu^2_2(\ds\frac{1}{3}|y_1-y_2|)+\nu^2_0(\frac{1}{3}|y_1-y_2|)
t^{-4}\right)
+\ds
16b^4
\sum\limits_{i,j=0,2}\frac{1}{t^i}\nu^2_{2-i}(|y_1-y_3|)
\cdot \frac{1}{t^j}\nu^2_{2-j}(|y_2-y_4|).
$$
The proof for the case $y_3\in I_3,$
$ y_4\in I_1$ is the same.
 \hfill$\Box$\\
~\\
{\bf Remark}
For a
translation-invariant measure $\mu_0$
the estimate similar to (\ref{m1}) is obtained  in
 \ci[inequality (20.42)]{Bi}.\\
~\\
{\it Step 2}
(\ref{m1})
holds with any permutations of $y_1,y_2,y_3,y_4$ in
the RHS. Hence
\beqn
&&|m_0^{(4)}(\bar y)|\le\ds
4b^4\left(
\nu^2_2(\frac{1}{3}|y_s-y_p|)+\nu^2_0(\frac{1}{3}|y_s-y_p|)
t^{-4}\right)\nonumber\\
&&+16b^4
\sum\limits_{i,j=0,2}
\left(\ds\frac{1}{t^i}\nu^2_{2-i}(|y_s-y_k|)
\cdot \frac{1}{t^j}\nu^2_{2-j}(|y_p-y_l|)
+
\frac{1}{t^i}\nu^2_{2-i}(|y_s-y_l|)
\cdot \frac{1}{t^j}\nu^2_{2-j}(|y_p-y_k|)\right)\nonumber\\
&&\equiv M^1_{s,p}(\bar y)+ M^2_{s,p}(\bar y)
\la{m2}
\eeqn
for any permutation $\{s,p,k,l\}$ of $\{1,2,3,4\}$.
Let us define
$$
\Sigma_{s,p}:=\{\bar z\in \Sigma^4~|~
|z_s-z_p|=\max_{i,j}|z_i-z_j|\}.
$$
Then
 $(\Sigma)^4=\bigcup\limits_{(s,p)}\Sigma_{s,p}$,
where the union is taken over all the pairs $(s,p)$
 of the indexes $1,2,3,4$.
Therefore, (\ref{m2}) implies
\be\la{115}
I(\bar x)\equiv
\int\limits_{\Sigma^{4}}
|m_0^{(4)}(\bar x-\bar z)|
dS(\bar z)\le
\sum\limits_{(s,p)}\{
\int\limits_{\Sigma_{s,p}}
 M^1_{s,p}(\bar x-\bar z)dS(\bar z)
+\int\limits_{\Sigma_{s,p}}
 M^2_{s,p}(\bar x-\bar z)dS(\bar z)\}.
\ee
Here the sum  is taken over all the pairs $(s,p)$.
Every of the six  terms
corresponding to different pairs $(s,p)$ in the RHS of (\ref{115})
coincide.
We have
to estimate
$I(\bar x)$ only for  $\bar x\in B_{r_0}^4$ (see (\ref{ix})).
Then
$|z_s-z_p-x_s+x_p|\ge (|z_s-z_p|-2{r_0})_+$
for any $ z_{s}, z_{p}\in \R^3$.
Since $\nu_d$ is a non-increasing function, (\ref{m2}),
 (\ref{115}) imply
\beqn
&&I(\bar x)\le
Cb^4
\int\limits_{\Sigma_{1,2}}
\left(
\nu^2_2(\frac{1}{3}(|z_1-z_2|-2{r_0})_+)+\nu^2_0(\frac{1}{3}(|z_1-z_2|-
2{r_0})_+)t^{-4}\right)dS(\bar z)
\nonumber\\
&&+Cb^4
\sum\limits_{i,j=0,2}\,\,\,
\int\limits_{\Sigma_{1,2}}
\left(\frac{1}{t^i}\nu^2_{2-i}((|z_1-z_3|-2{r_0})_+)
\cdot \frac{1}{t^j}\nu^2_{2-j}((|z_2-z_4|-2{r_0})_+)
\right)dS(\bar z)
\nonumber\\
&&\equiv I_1+I_2.\la{m4}
\eeqn
{\it Step 3}
Let us estimate $I_1$  and $I_2$ separately.
\begin{lemma}\la{I1}
$
I_1\le  C b^4|\Sigma|^2.
$
\end{lemma}
{\bf Proof}
The integrand in $I_1$
does not depend on $z_3$ and $z_4$.
Therefore,
the result of the integration in the $z_3, z_4$
we estimate by the  factor
$\pi|\Sigma||z_1-z_2|^2$, since
$|z_3-z_4|\le|z_1-z_2|$
by the definition of  $\Sigma_{1,2}$.
Lemma \ref{linb} implies
\beqn
I_1
&\le& C_1b^4|\Sigma|
\int\limits_{(\Sigma)^2} \Bigl(\nu^2_{2}
(\frac{1}{3}(|z_1-z_2|-2{r_0})_+)
+t^{-4}\nu^2_0(\frac{1}{3}(|z_1-z_2|-2{r_0})_+)\Bigr)
|z_1-z_2|^2
\,dS(z_1)dS(z_2)
\nonumber\\
&\le& C_1b^4
|\Sigma|^2
\int\limits_{0}^{2t}
\left(\nu_2^2(\frac{1}{3}(r-2{r_0})_+)+t^{-4}
\nu_0^2(\frac{1}{3}(r-2{r_0})_+)\right) r^3\,dr.\la{1111}
\eeqn
(\re{phi}) implies
$$
r^2\nu_2(r)=\nu_2(r)2\int\limits_0^r
sds\le 2\int\limits_0^r
s\nu_2(s)ds\le C<\infty.
$$
Therefore, using (\re{phi})  again,
\be\la{1112}
\int\limits_{0}^{2t} r^3
\nu_2^2(\frac{1}{3}(r-2{r_0})_+)\,dr\le C<\infty.
\ee
Finally, the integral
$$
\int\limits_{0}^{2t}
\nu^2_{0}(\frac{1}{3}(r-2{r_0})_+) \frac{1}{t^4}r^3\,dr=
\int\limits_{0}^{2t}
\frac{\nu_{0}(\ds\frac{1}{3}(r-2{r_0})_+)}{r}~
\frac{\nu_0(\ds\frac{1}{3}(r-2{r_0})_+)~r^4}{t^4}\,\,dr
$$
is bounded: it
converges to zero as $t\to\infty$
by Lebesgue theorem as in Lemma \ref{lgibs}.
Hence, Lemma \ref{I1} follows from (\ref{1111})  and (\ref{1112}).
\hfill$\Box$

\begin{lemma}\la{I2}
$
I_2\le  C b^4|\Sigma|^2.
$
\end{lemma}
{\bf Proof}
Since $\nu_d^2(r)\le C \nu_d(r)$, and $\Sigma_{1,2}\subset
\Sigma^4$, we have by
Lemma \ref{linb}
\beqn
\!\!\!\!\!\!\!\!\!\!\!&&I_2=C
b^4
\sum\limits_{i,j=0,2}\int\limits_{\Sigma_{1,2}}
\frac{1}{t^i}\nu_{2-i}(|z_1-z_3|-2{r_0})_+)
\cdot
\frac{1}{t^j}\nu_{2-j}(|z_2-z_4|-2{r_0})_+)
\,dS(\bar z)
\nonumber\\
\!\!\!\!\!\!\!\!\!\!\!
&&\le C b^4\sum\limits_{i,j=0,2}\,\,\,
\int\limits_{\Sigma^{2}}
\frac{1}{t^i}\nu_{2-i}(|z_1\!-\!z_3|\!-\!2{r_0})_+)
\,dS(z_1)dS(z_3)
\int\limits_{\Sigma^{2}}\frac{1}{t^j}\nu_{2-j}
(|z_2\!-\!z_4|\!-\!2{r_0})_+)
\,dS(z_2)dS(z_4)
\nonumber\\
\!\!\!\!\!\!\!\!\!\!\!
&&\le Cb^4 \sum\limits_{i,j=0,2}
|\Sigma|\int\limits_0^{2t}
\frac{r}{t^i}\nu_{2-i}((r-2{r_0})_+)\,dr \cdot
| \Sigma|\int\limits_0^{2t}\frac{r}{t^j}\nu_{2-j}
((r-2{r_0})_+)\,dr
\le Cb^4 |\Sigma|^2.\la{117}
\eeqn
In the last inequality we use
 (\ref{phi})
  and the Lebesgue theorem as
in Lemma \ref{lgibs}.
Lemma \ref{I2} is proved.
\hfill$\Box$
\bigskip\\
Now Lemmas \ref{I1}, \ref{I2} and (\ref{m4})
imply (\ref{ix}).
\hfill$\Box$



\setcounter{equation}{0}
\section{Convergence to equilibrium for variable coefficients}
We extend all results of previous sections to the case
of the wave equations with variable coefficients.
 We consider the wave equations in $\R^3$ with the
initial conditions
\beqn
\left\{\ba{l}
\ddot u(x,\,t)  =  \sum\limits_{j,k=1}^3
\pa_j(a_{jk}(x)\pa_k u(x,\,t))
- a_0(x)\,u(x,\,t),\,\,\,\,
x \in\R^3,\,\,t\in\R,\\
u|_{t=0} = u_0(x),~~\dot u|_{t=0} = v_0(x),
\ea
\right.\la{v1.1}
\eeqn
where $\ds\pa_j\equiv \fr\pa{\pa x_j}$.
We assume the following
properties {\bf E1--E3} of Eqn (\re{v1.1}).
\medskip\\
{\bf E1} $a_{jk}(x) = \delta _{jk} +  b_{jk}(x)$,
 where $b_{jk}(x) \in D$; also $a_0(x)\in D$.
\medskip\\
{\bf E2} $a_0(x)\geq 0$, and the
hyperbolicity condition holds: $\exists\al>0$ s.t.
\be\la{v1.3}
H(x, k )\equiv
\fr 12 \sum_{i,j=1}^3 a_{ij}(x) k_i  k_j\geq \al | k|^2,
\,\,\,\,x, k \in \R^3.
\ee
{\bf E3} Non-trapping condition holds, \ci{V89}:
for $(x(0), k(0))\in\R^3\times\R^3$ with $ k(0)\neq 0$
\be
\la{v1.4}
\vert x(t)\vert \rightarrow \infty,~~t\rightarrow \infty,
\ee
where $(x(t), k(t))$ is a solution to the following
Hamiltonian system
$$
\dot x(t)=\nabla_k H(x(t), k(t)),~~
\dot  k(t)=
-\nabla_x H(x(t), k(t)).
$$
{\bf Example}. {\bf E1}-{\bf E3} hold in the
case of constant coefficients,
$a_{jk}(x)\equiv\de_{ij}$.
For instance, {\bf E3}
hold because $\dot k(t)\equiv 0\Rightarrow
x(t)\equiv k(0)t + x(0)$.
\medskip

We denote as above,
$Y(t)\equiv (u(\cdot,t),\dot u(\cdot,t))$,
$Y_0\equiv (u_0,v_0)$.
Then  (\ref{v1.1}) becomes
\be\la{v1'}
\dot Y(t)={\cal F_*}(Y(t)),\,\,\,t\in\R,\,\,\,\,Y(0)=Y_0.
\ee

Proposition \re{p1.1} holds for the solutions to the
Cauchy problem  (\re{v1'}) as well as for  (\re{1'}).
Let $Y_0 $ in (\re{v1'})
be a measurable
random function with values in $({\cal H},\,{\cal B}({\cal H}))$,
and  let  $\mu_0$ be its distribution, as above.
Denote by $\mu_t$
the distribution of the solution $Y(t)$ to the problem
(\re{v1'}).
Let us state the extension of main Theorem \re{t1.2}.
We introduce the  appropriate Hilbert  spaces
of initial data of the infinite
energy. Let $\de$ be an arbitrary positive number.
\begin{definition}\la{vd6.1}
${\cal H}_{\delta}$
is the  Hilbert space of  functions $Y=(u,v)\in  {\cal H}$
with the finite norm
$$
\brr Y\brr^2_{\delta}=\int \ds e^{-2\delta|x|}
(|u(x)|^2+|\nabla u(x)|^2+|v(x)|^2)\,dx<\infty.
$$
\end{definition}

\begin{theorem}\la{A}
Let {\bf E1--E3},
{\bf S0--S3} hold.  Then \\
i) the convergence (\re{1.8}) holds for
any $\ve>0$.\\
ii)~The limit measure
 $ \mu_\infty $ is a Gaussian
measure on ${\cal H}$.\\
iii)~The limit characteristic functional has the form
$$
\ds\hat { \mu}_\infty (\psi ) = \exp
\{-\fr{1}{2} {\cal Q}_\infty (W \Psi,W \Psi)\},\,\,\,
\Psi \in {\cal D},
$$
where $W:\,{\cal D}\rightarrow {\cal H}_\de^\pr$
is a linear continuous operator for sufficiently small $\de>0$.
\end{theorem}

Theorem \re{A} follows immediately from Theorem \re{t1.2},
using the method \ci{DKR}. The method
is based on the scattering theory for the solutions of infinite
energy.


\setcounter{equation}{0}
\section{Appendix A. Radon transform}
{\bf Proof of Lemma \ref{l3.3}}
 Since
 $\ds\int\limits_{p\cdot\omega=  z\cdot\omega}
\!\!\!\!f(p) d^2p$
is an even function
with respect to $\omega$,
it suffices to prove the   next lemma.

\begin{lemma}\la{l5.1}
Let (\ref{phi}) hold, and $f\in C_\nu(\R^3)$.
Then

\be\la{4.1}
\frac{1}{(4\pi)^2}
\int\limits_{|\omega|=  1}\!\!dS(\omega)
\int\limits_{p\cdot\omega=  z\cdot\omega} f(p) d^2p=
-\frac{1}{2}{\cal E} *f(z),~~\forall z\in\R^3.
\ee
\end{lemma}
{\bf Proof}.
Both sides of (\ref{4.1})
define the continuous operators
$C_\nu(\R^3)\mapsto C_b(\R^3)$.
Therefore,
it suffices to consider
$f\in D$.
Applying the Fourier transform, we obtain with $\rho= | k|$,
\be\la{4.3}
({\cal E} *f)(z)=
\frac{1}{(2\pi)^3}
\int \hat {\cal E}( k)\hat
f( k)\ds e^{- i z\cdot  k}
d^3 k=
\frac{1}{(2\pi)^3}
\int\limits_{|\omega|=  1}\!\!dS(\omega)
\int\limits_0^{+\infty}\rho^2
\ds e^{- i\rho z\cdot\omega}
 \hat{\cal E}(\rho\omega)\hat f(\rho\omega)
 d\rho.
\ee
We substitute  $\ds\hat{\cal E}(\rho\om)=  -\fr1{ \rho^2}$
in the right hand side of (\ref{4.3}) and get
\be\la{4.4}
({\cal E} *f)(z)=
-\frac{1}{(2\pi)^3}\int\limits_{|\omega|=  1}\!\!dS(\omega)
\int\limits_0^{+\infty}
\ds e^{- i\rho z\cdot\omega}
\hat f(\rho\omega) d\rho.
\ee
Note that
\be\la{4.5}
\hat f(\rho\omega)=  \int\limits_{-\infty}^{+\infty}\
\ds e^{ i\rho  h}f^{\sharp}(h,\omega) dh,\quad
\mbox{where }
f^{\sharp}(h,\omega)\equiv
\int\limits_{y\cdot\omega=  h}f(y) d^2y.
\ee
Then from (\ref{4.4}), (\ref{4.5})
 we have
$$
({\cal E} *f)(z)=
-\frac{1}{(2\pi)^3}\int\limits_{|\omega|=  1}\!\!dS(\omega)
\frac{1}{2}\int\limits_{-\infty}^{+\infty}
\ds e^{- i y z\cdot\omega} dy
\int\limits_{-\infty}^{+\infty}
\ds e^{ i\rho h}f^{\sharp}(h,\omega) dh
$$
$$
=  -\frac{1}{8\pi^2}
\int\limits_{|\omega|=  1}dS(\omega)
F^{-1}_{y\to(z\cdot\omega)} F_{h\to y}
f^{\sharp}(h,\omega)=
-\frac{1}{8\pi^2}
\int\limits_{|\omega|=  1}
f^{\sharp}(z\cdot\omega,\omega) dS(\omega).
$$
Lemma \ref{l5.1} is proved.
\hfill$\Box$

\bigskip
{\bf Proof of Lemma \ref{l3.3'}}.
Since
$
F[P]( k)= - \ds\frac{i}{ | k|}\sgn k_3$,  we have
$$
(P *f)(z)=
\frac{1}{(2\pi)^3}
\int \hat P( k)\hat
f( k)\ds e^{- i z\cdot  k}
d^3 k=
-\frac{i}{(2\pi)^3}
\int\limits_{|\omega|=  1}\!\!dS(\omega)
\int\limits_0^{+\infty}\rho
\ds e^{- i\rho z\cdot\omega}
 \sgn(\omega_3)\hat f(\rho\omega)
 d\rho
$$
\be\la{4.6}
= - \frac{i}{(2\pi)^3}
\int\limits_{|\omega|=  1,\omega_3>0}
\!\!\!\!\!\!dS(\omega)
\int\limits_0^{+\infty}\rho
\ds e^{- i\rho z\cdot\omega}
\hat f(\rho\omega)
 d\rho+
\frac{i}{(2\pi)^3}
\int\limits_{|\omega|=  1,\omega_3<0}
\!\!\!\!\!\!dS(\omega)
\int\limits_0^{+\infty}\rho
\ds e^{- i\rho z\cdot\omega}
\hat f(\rho\omega)
 d\rho.
\ee
In the
last integral  we change the variables
 $\omega\to -\omega$, $\rho\to -\rho$, then
 apply (\ref{4.5})  and get
$$
(P *f)(z)= - \frac{i}{(2\pi)^3}
\int\limits_{|\omega|=  1,\omega_3>0}
\!\!\!\!\!dS(\omega)\int\limits_{-\infty}^{+\infty}
\rho \ds e^{- i \rho z\cdot\omega}
\hat f(\rho\omega)
 d\rho
$$
\be\la{4.7}
= - \frac{i}{(2\pi)^3}
\int\limits_{|\omega|=  1,\omega_3>0}
\!\!\!\!\! dS(\omega)
\int\limits_{-\infty}^{+\infty}
\ds e^{- i \rho z\cdot\omega}\,\,
\rho d\rho\int\limits_{-\infty}^{+\infty}
\ds e^{ i \rho h}
f^{\sharp}(h,\omega) dh.
\ee
Note that
\be\la{4.8}
\rho\int\limits_{-\infty}^{+\infty}
\ds e^{ i \rho h}
f^{\sharp}(h,\omega) dh=
 i \int\limits_{-\infty}^{+\infty}
\ds e^{ i \rho h}
(\nabla f)^{\sharp}(h,\omega)\cdot\omega\,dh,\,\,\,\rho\in\R.
\ee
Indeed, applying (\ref{4.5})
in the both sides of
$$
F[\nabla f](\rho\omega)\cdot\omega=  - i \rho
F[ f](\rho\omega),
$$
we obtain (\ref{4.8}).
Finally, from  (\ref{4.8}) and (\ref{4.7})
 we get

\beqn
(P *f)(z)&= &\frac{1}{(2\pi)^3}
\int\limits_{|\omega|=  1,\omega_3>0}
\!\!\!\!\! dS(\omega)
\int\limits_{-\infty}^{+\infty}
\ds e^{- i \rho z\cdot\omega} d\rho
\int\limits_{-\infty}^{+\infty}
\ds e^{ i \rho h}
(\nabla f)^{\sharp}(h,\omega)\cdot\omega\,dh\nonumber\\
~&&\nonumber\\
&= & \frac{1}{4\pi^2}
\int\limits_{|\omega|=  1,\omega_3>0}\!\!\!\!\!\!
F^{-1}_{\rho\to z\cdot\omega} F_{h\to \rho}
(\nabla f)^{\sharp}(h,\omega)\cdot\omega\, dS(\omega)\nonumber\\
~&&\nonumber\\
&= &
\frac{1}{4\pi^2}
\int\limits_{|\omega|=  1,\omega_3>0}
(\nabla f)^{\sharp}(z\cdot\omega,\omega)\cdot\omega\,dS(\omega)\nonumber\\
~&&\nonumber\\
&= &4{\cal P}f(z).
\eeqn

Lemma \ref{l3.3'} is proved.
\hfill$\Box$

\setcounter{equation}{0}
\section{Appendix B. Gaussian measures in Sobolev's spaces}

We verify (\re{Min}).
Definition (\re{ws}) implies for
$u\in{ H}_{s,\al}$,
\be\la{wsf}
\Vert u\Vert_{s,\al}^2=
\int(1+|x|)^{2\al}
\Big(
\int \ds e^{-ix( k-\eta)}
(1+| k|)^s(1+|\eta|)^s
\hat u( k)\ov{\hat u}(\eta)\,d k\, d\eta
\Big)dx.
\ee
Let $\mu(du)$ be a Gaussian translation invariant measure
in ${ H}_{s,\al}$ with a correlation
function $Q(x,y)=q(x-y)$.
Let us
introduce the following
correlation function
\be\la{corfu}
C( k,\eta)\equiv\int
\hat u( k)\ov{\hat u}(\eta)
\mu(du)
\ee
in the sense of distributions.
Since $u(x)$ is real-valued, we get
\be\la{corfuex}
C( k,\eta)
=F_{x\to k}F_{y\to-\eta}Q(x,y)=
C_n\de( k-\eta)\hat q( k).
\ee
Then,
integrating (\re{wsf}) with respect to the measure
$\mu(du)$, we get the formula
\be\la{wsmu}
\int
\Vert u\Vert_{s,\al}^2
\mu(du)=C_n
\int(1+|x|)^{2\al}dx
\int
(1+| k|)^{2s} \hat q( k)\,
d k.
\ee
Substituting $ \hat q( k)=1$ and 
$ \hat q( k)=| k|^{-2}$,
we get (\re{Min}).\hfill$\Box$


\end{document}